\documentclass[twocolumn]{aastex631}
\usepackage[all]{hypcap}
\usepackage[utf8]{inputenc}
\usepackage{amsmath}
\usepackage{booktabs}
\usepackage{xcolor}
\usepackage{pagecolor}
\newcommand{\tg}{t_{\gamma_d}}

\begin{document}

\defcitealias{Kaur_2025}{KS25}

\title{Binary disruptions driven by massive disks around massive black holes}
\author[0000-0002-3352-9272]{Mark Dodici}
\affiliation{Department of Astronomy \& Astrophysics, University of Toronto, Toronto, ON M5S 3H4, Canada}
\email{mark.dodici@astro.utoronto.ca}
\affiliation{Canadian Institute for Theoretical Astrophysics, Toronto, ON M5S 3H8, Canada}
\shorttitle{AGN disks cause binary disruptions}
\shortauthors{Dodici} 
\date{\today}
 
\begin{abstract}
    Accretion disks around massive black holes (MBHs) in galactic nuclei can contain significant mass, in which case their non-spherical potentials exert significant torques on single and binary stars orbiting the MBHs. 
    These torques can drive orbits to very large eccentricities.
    Previous works have shown that this driving can cause disruptions of single stars by the tidal gravity of the MBH. 
    Here, we characterize the ability of these torques to drive binary stars to the point of disruption via the Hills mechanism. 
    We derive semi-analytical estimates, validated with numerical simulations, of the number of binary disruptions driven by a generic disk in a generic nucleus. 
    Using these results, we estimate that the formation of the $\sim 5$-Myr-old disk of stars in the Galactic Center drove a burst of $\sim 10^2$ binary disruptions.
    These disruptions produced an excess of S-cluster stars and hypervelocity stars --- possibly including S5-HVS1, the fastest-known hypervelocity star. 
    In other galaxies, analogues of S-cluster stars produced by this process may evolve into tidal disruption events following a disk phase. 
    Hypervelocity star observations may help clarify the importance of disk-driven disruptions, which eject stars nearly isotropically and in temporal bursts --- a unique combination among processes driving binary disruptions.
\end{abstract}

\section{Introduction}
 
When a stellar binary approaches a massive black hole (MBH), its members may be separated by the hole's tidal force. When its orbit around the MBH, prior to disruption, had near-unity eccentricity, this process is known as the Hills mechanism \citep{Hills_1988}, and disruption yields two intriguing classes of stars. One binary member will be implanted on a short period orbit around the MBH, and the other will generally be ejected from the vicinity of the MBH at $\gtrsim 10^3$ km s$^{-1}$. In our Galaxy, the former may be observable as a member of the S-star cluster in the central few hundredths of a parsec around Sgr A* (e.g., \citealp{Gould_2003,Generozov_2020,Verberne_2025}; for observational review, see \citealp{Genzel_2010}). The latter may be observable as a hypervelocity star (HVS) in the Galactic halo (e.g., \citealp{Yu_2003,Brown_2005}; for a review, see \citealp{Brown_2015}).

Here, we describe how massive accretion disks drive a fraction of the binaries orbiting an MBH to become disrupted via the Hills mechanism. The result of this process is a burst of binary disruptions in a short window of time --- of order a few million years or less --- coincident with the arrival of the disk.
We refer to this process as ``accretion disk-driven disruption,'' or ADDD. This process should act in addition to ``standard'' disruptions driven by two-body relaxation \citep{Frank_1976,Lightman_1977,Yu_2003} or non-spherical structure of the background potential \citep{Magorrian_1999,Vasiliev_2013,Penoyre_2025}.
We later estimate that the time-averaged rate of binary disruptions through ADDD may be competitive with these processes, though uncertainties in nuclear cluster properties and the frequency of disk phases make such estimates imprecise.

The physics behind ADDD was clarified in 
\citet[hereafter \citetalias{Kaur_2025}]{Kaur_2025} 
who studied how accretion disks boost the rates of tidal disruptions of single stars 
(previous considerations of disk potentials in galactic nuclei include \citealp{Vokrouhlicky_1998,Subr_2004,Karas_2007}). 
Briefly, the non-spherical potential of the accretion disk exerts a torque on the orbits of stars and binaries around the MBH (hereafter ``outer'' orbits). 
This torque reduces pericenter separations of some outer orbits by decreasing their angular momenta at constant energy (i.e., increasing their eccentricities). 
As shown by \citetalias{Kaur_2025}, pericenter reduction may cause many single stars to reach their tidal radii and become disrupted, resulting in a ``burst'' of TDEs during a disk phase. 

\citetalias{Kaur_2025} do not discuss that binaries orbiting the MBH are subject to the same torque. 
While most binaries are relatively short-lived in such environments \citep[e.g.,][]{Alexander_2014}, the order-unity binary fraction observed among young stars in the central parsec around Sgr A* \citep{Gautam_2024,Bentley_2026} suggests that there may remain a non-negligible binary population among older stars, too.
Since the byproducts of ADDD are of observational interest, we aim to present a thorough characterization of disk-driven binary disruptions.

ADDD of binary stars has several observable implications. Primarily, it will produce many S-stars and HVSs in a short window of time. As there is a disk of stars in the central $0.5$ pc around Sgr A* that seems to have formed \emph{in situ} $\sim 5$ Myr ago \citep[e.g.,][]{Genzel_2003, Paumard_2006, Lu_2013,Yelda_2014, vonFellenberg_2022}, ADDD may have occurred recently in the Milky Way. 
In other galaxies, we argue that the S-star analogues produced by ADDD may evolve to yield (repeating/partial) tidal disruption events \citep[TDEs; e.g,][]{Bromley_2012,Rom_2026} in the $\sim$ 0.1--1 Gyr following an AGN phase. 

This work is structured as follows. 
First, we describe our setup (Section~\ref{sec:setup}) and characterize the orbits driven to disruption by a disk (Section~\ref{sec:orbits}), generalizing results of \citetalias{Kaur_2025}.
Then we discuss a model binary population and make semi-analytical estimates of the disrupted population from arbitrary clusters (Section~\ref{sec:estimates}), which we validate with numerical simulations (Section~\ref{sec:numerics}).
We estimate the number --- and observable byproducts --- of disruptions driven by the gas disk that likely formed the young stars around Sgr A* (Section~\ref{sec:CWD_impact}).
Finally, we discuss other possible implications of ADDD and the disk's torque, and we put ADDD in context with other mechanisms driving binary disruptions (Section~\ref{sec:discussion}).

\section{Potential and notation}\label{sec:setup}

\subsection{Notation}

\begin{figure} 
    \centering  
    \includegraphics[width=\linewidth]{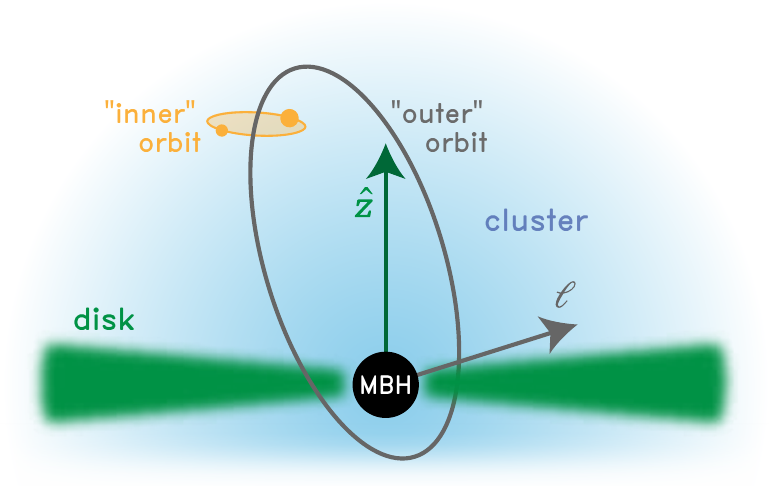}
    \caption{We consider a stellar binary orbiting an MBH. Its outer orbit is altered by the potential of a spherical cluster of background stars and a massive, axisymmetric accretion disk. The disk torques the outer orbit, changing its angular momentum $\ell$ --- and therefore its pericenter separation --- while conserving $\ell_z$, the component along the disk normal vector $\hat{z}$.} 
    \label{fig:cartoon}  
\end{figure}

Figure~\ref{fig:cartoon} shows our setup, now described in detail.
We choose a reference frame such that the $z$-axis coincides with the normal vector of the disk (i.e., the symmetry axis of the overall potential). The MBH sits at the origin; the distance from it is $r$.

Most orbits of interest are bound to the MBH, so it is natural to use the classical orbit elements, including
semimajor axis $a$, eccentricity $e$, inclination $i$, and argument of periapsis $\omega$.

We also use the ``reduced'' angular momentum and its $z$-component,
\begin{align}  \nonumber
    \ell
    = \sqrt{1-e^2}
    \quad \text{and} \quad \ell_z 
    = \ell\cos i.
\end{align}
When no disk is present, both $\ell$ and $\ell_z$ are conserved; under the axisymmetric potential of a disk, $\ell_z$ is still conserved, while $\ell$ is not.
As long as the potential is static, the semimajor axis is conserved. 

We will not use subscripts when referring to orbits about the MBH, which we will refer to as ``outer'' orbits. We refer to the motion of two binary members about their mutual center of mass as an ``inner'' orbit, and we distinguish elements of these orbits with subscript ``in.'' 

\subsection{Potential}

We say the NSC density $\rho_c$ is spherically symmetric and proportional to $r^{-\gamma_c}$. The enclosed cluster mass is equal to the MBH mass $m_\bullet$ at radius $r_h$ --- we call this the MBH radius of influence (though note there are other definitions of this term).

For the disk, we adopt the family of axisymmetric potential-density pairs presented by \citetalias{Kaur_2025}.\footnote{We confirmed that the potentials from several more-physical disk models yield long-term dynamics that are qualitatively equivalent to the \citetalias{Kaur_2025} disk family; see also Appendix~\ref{sec:appendix_pots}.} 
The potentials and densities of these disks are separable in spherical coordinates. Radially, the density $\rho_d$ is proportional to $r^{-\gamma_d}$. In polar angle, the density consists of a razor-thin midplane component --- containing most of the disk mass --- and a constant-scale-height envelope. The disk mass within $r_h$ is $m_d = \mu m_\bullet$, with $\mu$ a free parameter.

Explicit expressions for these potentials are included in Appendix~\ref{sec:appendix_pots}. 

From Appendix~\ref{sec:appendix_pots} and \citetalias{Kaur_2025}, the strength of the disk potential relative to the cluster potential is characterized by the 
parameter\footnote{The factor $(2-\gamma_c)/\alpha$ comes from a standard approximation to the orbit-averaged cluster potential; the parameter $\alpha$ (eq.~\ref{eq:alpha}) involves the gamma function, depends on $\gamma_c$ only, and varies monotonically from $0.5$ to $0$ as $\gamma_c$ goes from $1$ to $2$.}
\begin{equation}\label{eq:chi}
    \chi \equiv \mu \left(\frac{2-\gamma_c}{\alpha}\right) {\left(\frac{a}{r_h}\right)}^{\gamma_c - \gamma_d}.
\end{equation}

Typically, the disk is more centrally concentrated than the cluster, i.e., $\gamma_d > \gamma_c$. Then $\chi$ is largest --- the disk is at its most important --- at small $a$.
The opposite is true for $\gamma_d < \gamma_c$. When the slopes are equal, the relative strengths of disk and cluster do not change with $a$.

\begin{figure}
    \centering
    \includegraphics[width=\linewidth]{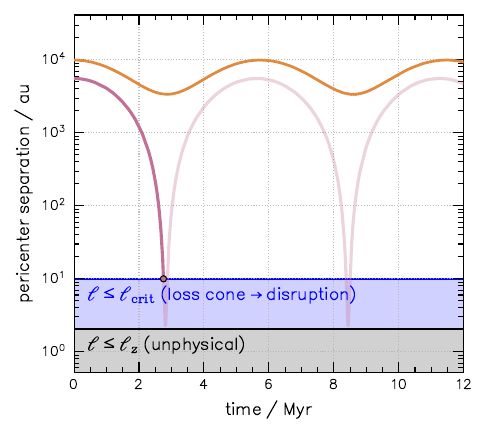}
    \caption{Two outer orbits evolve under a disk potential. In one (pink line), the binary on the orbit is driven to disruption --- its angular momentum reaches the critical value $\ell_{\rm crit}$ (eq.~\ref{eq:ell_crit}). The binary on the orange orbit is not driven to disruption. Semimajor axes are equal and constant, as is $\ell_z$. The only difference between the orbits is their initial $\ell$; the pink orbit is in the librating island, while the orange is in the upper circulating region (see Figure~\ref{fig:phase_space}). Simulations are described in Section~\ref{sec:numerics}.}
    \label{fig:examples}
\end{figure}

\subsection{Fiducial parameters}

We focus our examples on the Galactic Center, so we take $m_\bullet = 4\times10^6\,M_\odot$. We set $\gamma_c = 5/4$, motivated by observations of the Milky Way Nuclear Star Cluster (NSC; e.g., \citealp{Chatzopolous_2015,Gallego_Cano_2018,Schodel_2018}) suggesting the slope is shallower than the standard theoretical value ($\gamma_c = 7/4$; \citealp{Bahcall_1976}). (Outcomes are not particularly sensitive to this parameter; see also Fig.~5 of \citetalias{Kaur_2025}.) We set $r_h = 2\,{\rm pc}$.

We assume the cluster orbits have an isotropic velocity dispersion --- we set $\beta = 0$ in equation~(\ref{eq:DF}) (e.g., \citealp{Schodel_2009,Do_2013}; but cf.~\citealp{Feldmeier_2014}, who suggest a tangential bias).

Our fiducial disk has $\gamma_d = 5/2$ and $\mu = 0.1$. The slope is motivated by comparison to the \citet{Sirko_2003} and \citet{Thompson_2005} models of active galactic nucleus (AGN) disks (see Appendix~\ref{sec:appendix_pots}). The mass is motivated partially by these models --- in which the mass ratio may be as large as $\mu = 1$  --- and partially by the disk of young stars in the NSC --- assuming its present-day stellar mass $\sim 0.01m_\bullet$ is at least a few times smaller than the gas mass of its natal disk.

Note that the fiducial parameters have $\gamma_d > \gamma_c$; this inequality is relevant at several points.

\section{Outer orbits prone to disruption}
\label{sec:orbits}

\begin{figure}
    \centering 
    \includegraphics[width=\linewidth]{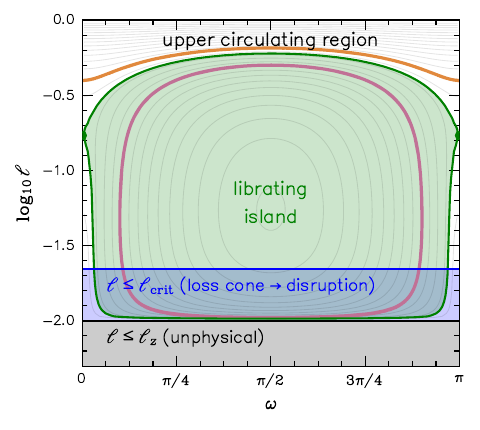}
    \caption{This phase portrait shows long-term orbit trajectories (grey contours) for orbits at fixed $\ell_z$. While all of these orbits are in the loss wedge, only some will reach the loss cone ($\ell \leq \ell_{\rm crit}$; blue) and be disrupted. 
    Specifically, most loss-wedge orbits in the librating island (green) or at smaller $\ell$ will reach the cone, while those in the circulating region above it will not. We show the evolution of the example orbits from Figure~\ref{fig:examples} as colored lines.}
    \label{fig:phase_space}
\end{figure}

In Figure~\ref{fig:examples}, we show the evolution of two orbits around a MBH under the cluster and disk potentials described in the previous section. The outer-orbit pericenter separations oscillate over time, as the orbits feel a torque from the axisymmetric disk potential. 
In one case, the pericenter separation becomes small enough for the binary to become tidally disrupted via the Hills mechanism.

How common should such disruptions be?
Briefly, an orbit may be torqued to the point of disruption if (i) it has very small $\ell_z$ and (ii) it has small enough initial $\ell$ that the disk causes its argument of pericenter $\omega$ to librate rather than circulate \citepalias{Kaur_2025}.

In this section, we summarize key results of \citetalias{Kaur_2025} in quantifying these conditions. We also present a novel fitting formula generalizing one of their results to arbitrary $\gamma_d$ and $\gamma_c$. 
This section yields a proportionality for the fraction of binaries disrupted by ADDD as a function of separation from the MBH (eq.~\ref{eq:fd_a}).

\subsection{Loss wedge of axisymmetric potentials}

A binary may be disrupted by the MBH if it reaches the tidal radius
\begin{equation}\label{eq:rt}
    r_t = a_{\rm in}{\left(\frac{m_\bullet}{m_b}\right)}^{1/3},
\end{equation}
with $m_b$ the binary mass. 
Disruption is not guaranteed at $r \lesssim r_t$ and may occur at $r \gtrsim r_t$ \citep[e.g.,][]{Sari_2010,Kobayashi_2012,Sersante_2025}, but this level of nuance is beyond the scope of this work.

An orbit's pericenter separation is within the tidal radius if its reduced angular momentum is smaller than a critical value, $\ell \leq \ell_{\rm crit}$. This condition defines the loss cone \citep[e.g.,][]{Lightman_1977}. For $a \gg r_t$, 
\begin{equation}\label{eq:ell_crit}
    \ell^2_{\rm crit} \simeq \frac{2r_t}{a}.
\end{equation}
For typical eccentricity distributions, the fraction of orbits in the loss cone is small.

Under a generic axisymmetric potential, $\ell$ may vary down to values as small as $\ell_z$ (the component of angular momentum along the symmetry axis), which is conserved. Then orbits with $\ell_z \leq \ell_{\rm crit}$ may enter the loss cone. This condition encloses the standard loss \emph{wedge} for axisymmetric potentials \citep[e.g.,][]{Magorrian_1999,Vasiliev_2013}, which encompasses a much larger volume of parameter space than the loss cone.

\subsection{Librating island}

Not all loss-wedge orbits will reach the loss cone, though. In order to do so, the disk torque must induce significant change in $\ell$ before the background cluster induces significant apsidal precession. \citetalias{Kaur_2025} defined the ``librating island'' of phase space, which hosts the vast majority of orbits that meet this criterion. It is only orbits in this island (akin to ``saucer'' orbits in generic axisymmetric potentials; see, e.g., Section 4.4.2 of \citealp{Merritt_2013}), or those at lower $\ell$, that experience substantial oscillations in $\ell$.

To illustrate this island, in Figure~\ref{fig:phase_space}, we show contours of the combined, orbit-averaged\footnote{For orbits near $r_h$, the cluster mass will induce significant precession in one orbital period, so orbit-averaging is not entirely valid. We use an orbit-averaged phase portrait for building intuition and making analytical estimates, but our later numerical work involves no averaging.} 
potential of the disk and cluster (evaluated numerically, for our fiducial parameters, at $a=0.1r_h$ and $\ell_z = 10^{-2}$). The librating island is delineated by a separatrix (green contour) and shaded in green. 
\citetalias{Kaur_2025} show many more examples of librating islands under various conditions.

On a timescale much longer than its period (roughly eq.~\ref{eq:t_sec}), an orbit should evolve along a contour of this phase portrait.
In this example, while every contour represents an orbit in the loss wedge ($\ell_z \leq \ell_{\rm crit}$), 
only orbits in or below this island enter the loss cone ($\ell \leq \ell_{\rm crit}$; blue shaded region).

\citetalias{Kaur_2025} estimate that only orbits that may enter the librating island are driven to the loss cone. 
An orbit may enter the island if its initial action, $J$, is smaller than the action of the librating island separatrix, $J_{\rm sep}$, when the disk is at its most massive (Section 3, \citetalias{Kaur_2025}). 
If the maximum separatrix action is smaller than an orbit's initial action, that orbit will forever be confined to the upper circulating region --- its minor oscillations in $\ell$ are unlikely to bring it to the loss cone, so it should avoid disruption.

Using this condition effectively requires a good grasp on the separatrix action, $J_{\rm sep}$, associated with different combinations of disk and cluster properties.

\subsection{Separatrix action}

\citetalias{Kaur_2025} find that the separatrix action is well fit by $J_{\rm sep} \propto \chi^{1/2}$, where $\chi$ (eq.~\ref{eq:chi}) is the relative disk strength for a given combination of $\mu$ and $a$. Their fit only holds for $\gamma_d = 3/2$ and $\gamma_d < \gamma_c$, though. 
Through numerical evaluation of the orbit-averaged potential, we find that the separatrix action is always well fit by 
\begin{equation}\label{eq:jhyp_fit}
    J_{\rm sep} = J_{\rm sat}\left(\frac{\chi/\chi_{\rm sat}}{1+\chi/\chi_{\rm sat}}\right)^{1/2}.
\end{equation}
The parameters $J_{\rm sat}$ and $\chi_{\rm sat}$ are fit for a particular value of $\gamma_d$; we discuss fits in Appendix~\ref{sec:appendix_pots}. 

When $\chi \ll \chi_{\rm sat}$, the separatrix action grows like $\chi^{1/2}$; when $\chi/\chi_{\rm sat} \gtrsim 1$, the action saturates to $J_{\rm sep} \rightarrow J_{\rm sat}$. 
The orbits in \citetalias{Kaur_2025} are almost entirely in the $\chi \ll \chi_{\rm sat}$ regime, 
so equation~(\ref{eq:jhyp_fit}) yields their fitting formula.

\subsubsection{Disrupted fraction vs.~separation from the MBH}

With this generic fitting formula, we can estimate how the fraction of binaries subject to ADDD scales with outer-orbit semimajor axis.

When $\gamma_d > \gamma_c$, the separatrix action is saturated interior to some radius $r_{\rm sat}$ from the MBH. The value of $r_{\rm sat}$ can be found by setting $\chi=\chi_{\rm sat}$ and solving for $a$, which yields 
\begin{equation}\label{eq:rsat}
    \frac{r_{\rm sat}}{r_h} = {\left[\frac{\chi_{\rm sat}}{\mu}\left(\frac{\alpha}{2-\gamma_c}\right)\right]}^{1/(\gamma_c - \gamma_d)}.
\end{equation}
For our fiducial parameters, $r_{\rm sat}/r_h = 0.03$, that is, the separatrix action is saturated in the inner few hundredths of the radius of influence. For $\mu=0.01$ (a ten-times-less-massive disk), $r_{\rm sat}/r_h = 0.001$.

\begin{figure}
    \centering
    \includegraphics[width=\linewidth]{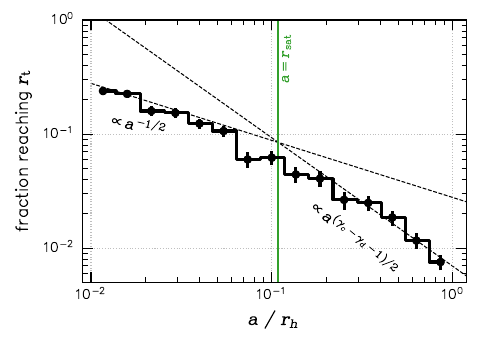}
    \caption{For fixed binary properties, the fraction of systems reaching the tidal disruption radius ($r_t$) is approximately a piecewise power law in distance from the MBH (eq.~\ref{eq:fd_a}). The break occurs at the separatrix saturation radius, $r_{\rm sat}$ (green line; eq.~\ref{eq:rsat}). Here, the black line shows results of numerical integrations of outer orbits; $r_t$ is set by $m_b=4\,M_\odot$ and $a_{\rm in}=1$ au. The dashed lines show the noted scalings, with arbitrary normalization.}
    \label{fig:frac_disrupted_vs_a}
\end{figure}

Interior to $r_{\rm sat}$, the fraction of loss-wedge orbits leading to disruptions will be roughly constant. Exterior to this radius, this fraction will decay as $a^{(\gamma_c - \gamma_d)/2}$. The size of the loss wedge is proportional to $a^{-1/2}$, so the fraction of orbits destroyed at a given $a$ is
\begin{equation}\label{eq:fd_a}
    f_{d}(a) \propto 
    \begin{cases}
        a^{-1/2}, \quad \quad \;\;\;\quad a \lesssim r_{\rm sat}
        \\ 
        a^{(\gamma_c - \gamma_d - 1)/2}, \quad a \gtrsim r_{\rm sat},
    \end{cases}
\end{equation}
assuming the $\ell$ distribution is uniform and that the $\ell_z$ distribution does not vary with $a$. This piecewise scaling is shown in Figure~\ref{fig:frac_disrupted_vs_a}, validated with simulation results.

When $\gamma_d < \gamma_c$, the separatrix action saturates exterior to $r_{\rm sat}$. In this regime, the saturation radius is almost always beyond the radius of influence.
(For the fiducial parameters in \citetalias{Kaur_2025} --- $\gamma_d = 3/2$, $\gamma_c = 7/4$, $\mu = 0.1$, for which $\chi_{\rm sat} = 1.8$ and $\alpha=0.09$ --- we find $r_{\rm sat}/r_h \sim 10^3$.) Then when $\gamma_d < \gamma_c$, the fraction of orbits destroyed at a given $a$ always has the latter scaling in equation~(\ref{eq:fd_a}).

\section{Disruptions from a coeval cluster} 
\label{sec:estimates}

We now have a sense of the parameters of an orbit that will be driven to disruption by an accretion disk --- it must be in the loss wedge, and it should be in or below the librating island.

On a population level, the efficacy of ADDD is primarily set by two cluster properties: the (an)isotropy of orbits 
and the age of the cluster at the time of disk growth. In this section, we show how to make estimates of the number of disrupted binaries, while discussing the importance of each of these properties and the construction of a physically motivated binary population.

We consider disruptions from a cluster of stars born simultaneously. Observational and theoretical arguments suggest NSCs are built up over time \citep[for a review, see][]{Neumayer_2020}, so such a coeval cluster should be imagined as a subpopulation of the broader NSC, comprised of $N_c$ stars (with $N_c \leq N_h$, the total number of stars within the radius of influence). 

\subsection{Estimates of disrupted population}

\citetalias{Kaur_2025} used the loss wedge and librating island to estimate the number of stellar tidal disruptions induced by a massive disk. They integrated the distribution function of orbits, $F(a,\ell)$, over the regions of parameter space encompassed by both criteria, considering a cluster comprised solely of single stars.

Similarly, the distribution of semimajor axes of disrupted binaries may be estimated as
\begin{equation}\label{eq:integral_basic}
    \left(\frac{dN_d}{da_{\rm in}}\right)_{m_b} = \int_{0}^{r_h} da \int_{-\ell_{\rm crit}}^{\ell_{\rm crit}}  d\ell_z \int_{0}^{J_{\rm sep}} d\ell \left(\frac{dF}{da_{\rm in}}\right),
\end{equation}
where $dF/da_{\rm in}$ is the differential contribution to the distribution function from binary stars of given $a_{\rm in}$. The bounds of the integral over $\ell$ encompass the librating island, and those of the integral over $\ell_z$ limit us to the loss wedge.
Note that equation~(\ref{eq:integral_basic}) is for a population at fixed binary mass, $m_b$.

This distribution is useful for estimating properties of hypervelocity or S-stars resulting from ADDD.
The total number of binary disruptions, $N_d$, comes from integrating equation~(\ref{eq:integral_basic}) over $m_b$ (after convolution with a mass function) and $a_{\rm in}$ --- i.e., 
\begin{equation}\label{eq:final_Nd}
    N_d = \int da_{\rm in} \int dm_b \left(\frac{1}{N_c}\frac{dN_c}{dm_b}\right)\left(\frac{dN_d}{da_{\rm in}}\right)_{m_b},
\end{equation}
with $N_c$ the number of orbits in the coeval cluster.

\subsection{The importance of (an)isotropy}
We assume that cluster orbits follow a double-power law distribution function \citep{Kaur_2018}
\begin{equation}\label{eq:DF}
    F(a,\ell) = (1-\beta)(3-\gamma_c)\frac{N_h}{r_h}{\left(\frac{a}{r_h}\right)}^{2-\gamma_c}\ell^{-2\beta},
\end{equation}
which is normalized such that its integral over $a\in[0,r_h]$, $\ell \in [0,1]$, and $\ell_z \in[-\ell,\ell]$ yields $N_h$, the number of orbits in the radius of influence. 

The parameter $\beta$ quantifies the velocity anisotropy of the cluster orbits \citep[e.g.,][]{BT_2008}. A thermal eccentricity distribution (flat in $e^2$; \citealp{Jeans_1919}) corresponds to $\beta = 0$, which is said to be an ``isotropic'' cluster. Clusters with $\beta < 0$ are said to be ``tangentially'' biased (in terms of $e$, these have a circular bias relative to thermal), and those with $\beta > 0$ are said to be ``radially'' biased (an eccentric bias relative to thermal).

Very eccentric orbits, which have very small initial actions $J$, are most likely to enter the librating island. Therefore a cluster with larger $\beta$ will have more disruptions (e.g., Figure 5 of \citetalias{Kaur_2025}).

\subsection{Binary population}\label{sec:binary_population}

The initial conditions --- and subsequent evolution --- of a binary population plays a crucial role in determining the number of disk-driven disruptions.
We use the following toy model for $dF/da_{\rm in}$:
\begin{enumerate}
    \item At the radius of influence, $a=r_h$, there is some binary fraction $f_{b,h}$.
    \item Binary semimajor axes at $a=r_h$ are sampled from a distribution $df_{b,h}/d a_{\rm in}$. The minimum value is set at $a_{\rm in,\min} = 2.5R_\odot(m_1/M_\odot)^{0.8}$ as an approximation for a contact binary;
    $m_1$ is the mass of the more-massive binary member, so this is roughly $2.5\times$ that member's radius. The maximum value is set by a tidal stability criterion, $a_{\rm in,\max} = 0.1r_h(m_b/m_\bullet)^{1/3}$ \citep[see][]{Grishin_2017,Vynatheya_2022}.
    \item At $a<r_h$, binary semimajor axes are sampled from the same distribution, with the same binary fraction. Some will be wider than the \emph{local} tidal stability criterion, i.e., $a(m_b/m_\bullet)^{1/3}$; these systems are discarded and treated instead as singles. 
    This rejection naturally provides a binary fraction that is smaller closer to the MBH.
    \item At each $a$, a fraction $\left[1-(a/r_h)^{\lambda}\right]$ of the remaining binaries are discarded. 
    Physically, when $\lambda = 0$, this model assumes that the efficacy of forming a binary at some $a_{\rm in}$ does not change with $a$, except for the limit imposed by the tidal force of the MBH. Setting $\lambda > 0$ implies that binary formation is less efficient at smaller $a$ and that this inefficiency does not depend on $a_{\rm in}$. 
    \item Before a disk phase, binaries may be destroyed through interactions with other stars after an evaporation timescale, $t_{\rm evap}$. This condition may be converted to a minimum separation from the MBH at which one finds binaries of given properties at a given cluster age; we call this separation $r_m(a_{\rm in}, m_b, t_{\rm age})$. We make the approximation that a binary's properties do not change until $t_{\rm age} = t_{\rm evap}$.
    \item Binaries in which the more-massive member has left the main sequence, after time $t_{\rm MS}(m_1)$, are removed from the population.
\end{enumerate}
We take $f_{b,h} = 1$ and $\lambda = 0$ throughout this work but retain the parameters in our discussion for thoroughness.

This model can be written quantitatively as
\begin{equation}\label{eq:dF_da}
    \frac{dF}{d a_{\rm in}} =  \left(\frac{df_{b,h}}{da_{\rm in}}\right){\left(\frac{a}{r_h}\right)}^{\lambda} F(a,\ell)
\end{equation}
for $a > r_m(t_{\rm age})$ and $t_{\rm MS}(m_1) > t_{\rm age}$; if either condition is not met, $dF/da_{\rm in} = 0$.
The minimum separation from the MBH at which a binary may exist, $r_m$, evolves with time; we discuss this evolution in the following subsection.
We assume a log-uniform distribution of $a_{\rm in}$, i.e.,
\begin{equation}
    \frac{df_{b,h}}{d a_{\rm in}} = \frac{f_{b,h}}{\ln(a_{{\rm in},\max,h}/a_{\rm in,\min})}\frac{1}{a_{\rm in}}.
\end{equation}

Observational constraints for our model are sparse, so we have chosen to construct a simple population with physical motivation (which conveniently yields a distribution function separable between $a_{\rm in}$, $a$, and $\ell$). In our own NSC, our strongest constraint is that the binary fraction seems to grow with separation from Sgr A* \citep{Chu_2023,Gautam_2024,Bentley_2026}, but it is unclear whether this feature is set by initial conditions or dynamical evolution of the binary population \citep{Stephan_2016,Dodici_2026}. Such a trend arises naturally in our model at all times (even when $\lambda = 0$). We have chosen a semimajor axis distribution following Öpik's law \citep{Opik_1924}, though the model is entirely flexible to other choices.

\subsection{The importance of age}
Over time, wide binaries are depleted through binary--single interactions. The result of this depletion is that a cluster that is older at the onset of a disk phase has fewer binaries available to undergo ADDD. 

In our model, this depletion is reflected in the growth of $r_m$ with cluster age.
At formation, $r_m$ is set by the tidal stability criterion, but at later times, it is set by the distance from the MBH where $t_{\rm evap} = t_{\rm age}$ for binaries of given $a_{\rm in}$ and $m_b$. 

The evaporation timescale is \citep[e.g.,][]{Alexander_2014}
\begin{equation}
    t_{\rm evap} = \frac{\sqrt{3}}{16\sqrt{\pi}}\left(\frac{m_b}{m_\star}\right) \frac{\sigma_v}{G \rho_c a_{\rm in} \ln \Lambda_c},
\end{equation}
where $m_\star$ is the average perturber mass, $\rho_c$ is the local cluster mass density, and $\ln \Lambda_c$ is the Coulomb logarithm, which we approximate as $\ln(2\sigma_v^2/v_{\rm kep}^2)$. 
We assume the velocity dispersion scales like $r^{-\gamma_v}$ and is
$\sigma_{v,h}$ at the radius of influence.

Equating $t_{\rm age}$ and $t_{\rm evap}$ 
and solving for $a$ 
shows that 
\begin{align}\label{eq:rm}
    \frac{r_m}{r_h}
    \propto {\left(\frac{a_{\rm in}t_{\rm age}}{m_b}\right)}^{1/(\gamma_c - \gamma_v)}
\end{align}
where we have used the power-law profiles\footnote{We assume $\gamma_c > \gamma_v$; if this inequality does not hold, the evaporation timescale is instead shortest at large $a$. The integration over $a$ in equation~(\ref{eq:integral_basic}) would then be bounded on the low end by the tidal limit and on the high end by $r_m$.} 
of $\sigma_v$ and $\rho_c$ and ignored the weak dependence of $\ln\Lambda$ on $a$.
Once $r_m = r_h$, there are no more binaries of a particular $a_{\rm in}$ and $m_b$ within the radius of influence. 

\begin{figure*}
    \centering
    \includegraphics[width=\linewidth]{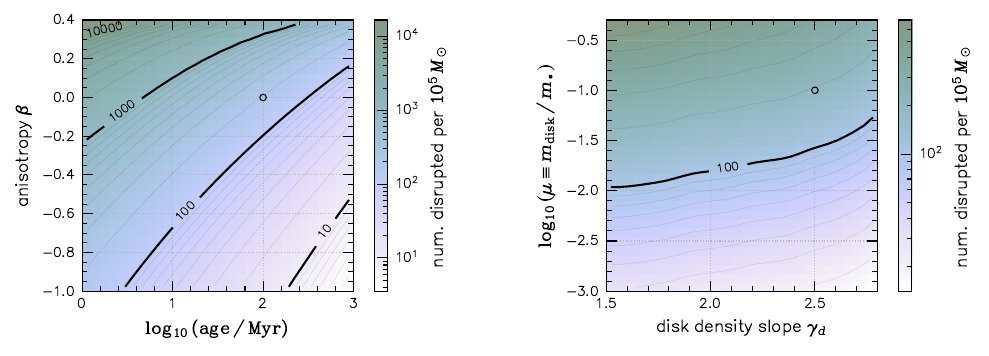}
    \caption{The number of disrupted binaries increases with cluster anisotropy $\beta$ and decreases with cluster age (left). The number increases with disk mass $\mu$ and decreases with disk concentration $\gamma_d$ (right), but less steeply than with cluster properties. The numbers shown are for a cluster of $10^5\,M_\odot$ of stars. 
    The small circle shows the fiducial parameters, which are also used in Figure~\ref{fig:compare_approaches} --- an isotropic cluster born 100 Myr before the disk arose.}
    \label{fig:num_vs_cluster}
\end{figure*} 

\subsection{Analytical estimates}

In Appendix~\ref{sec:appendix_analytic}, we find an approximate analytical solution to equation~(\ref{eq:integral_basic}). The disrupted-binary distribution is proportional to
\begin{align}\label{eq:dist_result}
   \left(\frac{dN_d}{da_{\rm in}}\right)_{m_b} 
   &\propto 
   J_{\rm sat}^{1-2\beta}
   {\left(\frac{r_h}{r_{\rm sat}}\right)}^{\Gamma}
   {\left(\frac{r_t}{r_h}\right)}^{1/2}
   \left(\frac{df_{b,h}}{da_{\rm in}}\right)\mathcal{F},
\end{align}
with $\Gamma \equiv (\gamma_c - \gamma_d)(1/2-\beta)$. The fitting factor $J_{\rm sat}$ is discussed around equation~(\ref{eq:jhyp_fit}), while $\mathcal{F}$ will be discussed shortly.

The first two terms in this expression have no dependence on $a_{\rm in}$; they relate to the size of the librating island throughout the cluster. Larger $J_{\rm sat}$ means more disruptions. If the disk is more concentrated than the cluster ($\gamma_d > \gamma_c$), smaller $r_{\rm sat}/r_h$ means fewer disruptions; if $\gamma_c > \gamma_d$, the opposite is true.
The disk mass enters through $r_h/r_{\rm sat} \propto \mu^{1/(\gamma_c - \gamma_d)}$, such that
\begin{equation}\label{eq:prop_mu}
 \left(dN_d/da_{\rm in}\right)_{m_b} \propto \mu^{1/2-\beta}.
\end{equation}
Therefore in an isotropic cluster ($\beta=0$), the efficacy of ADDD only scales like the square root of the disk mass.

The slope of the disrupted-binary semimajor axis distribution comes from $r_t^{1/2 }\propto a_{\rm in}^{1/2}$ and $df_{b,h}/da_{\rm in}$. The slope is therefore steeper than the primordial population distribution by a factor $a_{\rm in}^{1/2}$. 

The reduction factor $\mathcal{F}$, which depends on $r_m$ and thereby $a_{\rm in}$, is given explicitly in Appendix~\ref{sec:appendix_analytic}. It truncates the distribution of disrupted-binary semimajor axes --- in a cluster of some age, $\mathcal{F}$ causes the distribution to be $0$ at $a_{\rm in}$ values where all binaries in the cluster would have evaporated (i.e., where $r_m \geq r_h$).

\subsection{Influence of cluster and disk properties}\label{sec:trends}

We can use the results of this section to estimate the number of binaries disrupted from a coeval cluster of some total mass. 
In Figure~\ref{fig:num_vs_cluster}, we show how the total number of disrupted binaries varies with properties of the cluster and disk. For these figures, we evaluate equation~(\ref{eq:final_Nd}) numerically, using our analytical approximation for equation~(\ref{eq:integral_basic}), taking the coeval cluster to have mass $10^5\,M_\odot$. 

In solving equation~(\ref{eq:final_Nd}), we need to choose a binary mass distribution. We say the more-massive binary member, $m_1$, comes from an initial mass function (IMF) in which the number of stars between $m$ and $m + dm$ is proportional to $m^{-\alpha_m}$, with $\alpha_m = 2.3$ \citep{Salpeter_1955,Kroupa_2002}, in the range $m_1 \in [1,100]\,M_\odot$. For simplicity, we take all binary mass ratios to be equal to $\bar{q} = 0.55$, i.e., the mean value for a uniform distribution on $[0.1,1]$ (in Monte Carlo-sampled populations, discussed later, we actually draw from this distribution). The total number of binaries in the coeval population is then $N_c = m_c/\left[\bar{m}_1(1+\bar{q})\right]$, where $m_c$ is the total population mass and $\bar{m}_1(\alpha_m)$ is the mean primary mass.

From Figure~\ref{fig:num_vs_cluster}, the cluster anisotropy parameter and age are critical in determining the number of disruptions. As discussed, young, radially biased clusters are best for ADDD. While we do not show it here, the cluster concentration parameter $\gamma_c$ has very little influence on $N_d$ (see also \citetalias{Kaur_2025}). 

Dependence on disk properties is weaker, but still notable. A more-massive, less-concentrated disk yields more disruptions. The mass dependence is straightforward; a heavier disk makes the total potential more non-spherical. The concentration-dependence is sensible considering the limit $\gamma_d \rightarrow 3$: if all of the disk mass were internal to an orbit, its orbit-averaged potential would have no dependence on $\omega$, so the disk could not alter angular momentum $\ell$ (the disk mass would simply add a quadrupole moment to the MBH).  

\subsection{Monte Carlo sampling of population}\label{sec:monte_carlo}

The (semi-)analytical work done so far provides useful insight into the dependence of ADDD on various properties of the cluster and disk. That said, Monte Carlo solutions to integrals~(\ref{eq:integral_basic}) and (\ref{eq:final_Nd}) provide more flexibility and do not require any approximation.

For the remainder of this work, results come from Monte Carlo sampling of these integrals unless otherwise specified. 
These sampled populations follow the model in Section~\ref{sec:binary_population}.
The number of binaries disrupted by a disk arising at $t_{\rm age}$ after this population formed is simply the number (i) in both the loss wedge and librating island ($|\ell_z| \leq \ell_{\rm crit}$ and $\ell \leq J_{\rm sep}$), (ii) with a primary main-sequence lifetime $> t_{\rm age}$, and (iii) with an evaporation timescale $> t_{\rm age}$.
A comparison between the (semi-)analytical work above, this Monte Carlo sampling, and numerical integrations is shown in Figure~\ref{fig:compare_approaches}.

\section{Numerical validation}\label{sec:numerics}

To validate these estimates, we simulate test particles orbiting an MBH using the $N$-body integrator package \texttt{rebound} \citep{Rein_2012,Rein_2015}. We add the acceleration from the cluster and disk potentials using the custom-effect helper package \texttt{reboundx} \citep{Tamayo_2020}.

We simulate 22,132 orbits. We sample $a$ and $\ell$ from equation~(\ref{eq:DF}), with $\gamma_c = 1.25$, $\beta = 0$, and $i$ isotropic. We integrate for $10t_{\rm sec}$ (eq.~\ref{eq:t_sec}) --- ten times the precession timescale from the spherical cluster, alone --- recording the pericenter separation each orbit. We prescribe for each orbit some $a_{\rm in}$ and $m_b$ based on our binary population model, and we record whether or not it reaches the tidal disruption radius (eq.~\ref{eq:rt}). Two example orbits are shown in Figures~\ref{fig:examples} and \ref{fig:phase_space}.

\begin{figure} 
    \centering
    \includegraphics[width=\linewidth]{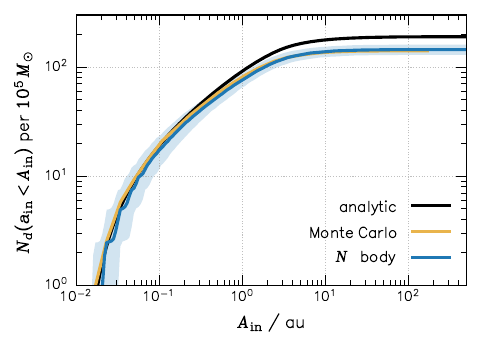}
    \caption{The distribution of semimajor axes of disrupted binaries from a coeval cluster is well approximated by an integral over the loss wedge and librating island (eq.~\ref{eq:integral_basic}). Here, we validate our analytical and Monte-Carlo sampled solutions to this integral against results of numerical simulations.}
    \label{fig:compare_approaches}
\end{figure}
In Figure~\ref{fig:compare_approaches}, we show the cumulative distribution of $N_d$ over $a_{\rm in}$ from several methods. We show results from analytical and Monte-Carlo evaluation of equation~(\ref{eq:integral_basic}), as discussed in the previous section, and results from our numerical simulations.

There is excellent agreement between the Monte-Carlo solution of equation~(\ref{eq:integral_basic}) and the outcomes of the numerical simulations. As expected (see Appendix~\ref{sec:appendix_analytic}), the analytical evaluation somewhat over-estimates the number of disruptions, but the error is not significant given the uncertainties in, e.g., cluster anisotropy and age.

It is worth noting that almost all disrupted binaries in this example have $a_{\rm in} < 1$ au. The widest binary disrupted by ADDD is roughly determined by the age of the cluster ($100$ Myr for this example); specifically, it is the widest binary whose evaporation timescale at the radius of influence ($a=r_h$) is longer than the cluster age. In the notation of the previous section, it is the widest binary for which $r_m < r_h$.

Through these simulations, we verified that many orbits ``jump'' into the loss cone from one pericenter passage to another, as a result of significant change in $\ell$ on timescales comparable to the outer-orbital period. As a result, many binaries are disrupted with pericenter separations much smaller than $r_t$ (i.e., $\ell \ll \ell_{\rm crit}$). 
Such deep plunges are reassuring for our assumption that any orbit reaching $r \leq r_t$ yields a disruption, as $r\ll r_t$ pericenter passages are much more likely to result in disruption \citep[e.g.,][]{Sersante_2025}.

\section{Recent disk in the Galactic Center}
\label{sec:CWD_impact}

Now that we have characterized ADDD, let us apply the results to a relevant example.

The orbits of most young (O-type) stars in the central half-parsec around Sgr A* are coplanar with each other and nearly face-on from our perspective, forming a structure known as the clockwise disk (CWD; \citealp{Genzel_2003,Paumard_2006,vonFellenberg_2022}).
This disk presents strong evidence of a star formation event $\sim 5$ Myr ago, in which stars apparently formed from a massive disk of gas \citep[e.g.,][]{Levin_2003,Nayakshin_2005,Nayakshin_2007,Bonnell_2008,Hobbs_2009}. 
Estimates of the total mass of the CWD are $\sim 0.01m_\bullet$. The gaseous disk from which it formed may have been more massive --- we estimate this gas disk had $\mu = 0.1$. 

We construct a Monte Carlo-sampled Milky Way NSC and estimate the number of disruptions driven by the disk that formed the CWD. We discuss the properties of the resultant S-stars and hypervelocity stars.

\subsection{Nuclear cluster star formation history}
The Milky Way NSC is well-studied, but its formation history is still uncertain. In the previous section, we saw that younger clusters beget more binary disruptions, so this formation history is important. 

We focus on three possible NSC star formation histories. In the first, most ($75\%$) of the mass formed $10$--$13$ Gyr ago, a sizeable fraction ($20\%$) formed from $2$--$4$ Gyr ago, and the remainder formed $100$--$500$ Myr ago (roughly based on \citealp{Schodel_2020,Gallego-Cano_2026}). In the second, all of the mass formed $5$ Gyr ago (roughly based on \citealp{ZChen_2023}). 
In the third, the NSC star formation rate was uniform in time from $5$ Myr to $13$ Gyr ago.

We sample the total population of the NSC following the Monte Carlo approach detailed in Section~\ref{sec:monte_carlo}, then prescribe ages based on these histories. 
For each model, we draw 30 instances of the NSC and report median values and $1\sigma$ ranges.

\subsection{ADDD in the CWD formation episode}\label{sec:ADDD_during_CWD}

In all three models, we assume a massive gas disk arose $5$ Myr ago. We set this disk mass to be $0.1m_\bullet$, i.e., we approximate that $10\%$ of the disk mass was converted to stellar mass in the present-day CWD. 

\begin{figure}
    \centering
    \includegraphics[width=\linewidth]{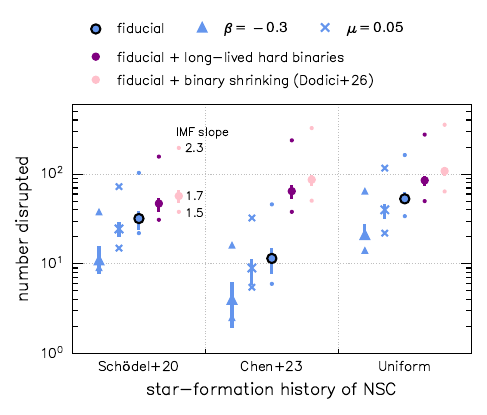}
    \caption{During the formation of the young disk of stars around Sgr A*, between $\sim 10$ and $\sim 10^2$ binaries should have been disrupted by ADDD.
    Each column shows estimates for a given NSC star formation history, with a variety of assumptions about cluster and disk properties (triangles, xs, and small dots) and improved aspects of pre-disk binary dynamics (different colors). 
    These variations are discussed in Section~\ref{sec:ADDD_during_CWD}.}
    \label{fig:number-disrupted_CWD}
\end{figure}

The number of disrupted binaries for each formation history is shown in Figure~\ref{fig:number-disrupted_CWD}.
Various assumptions lead to estimates of anywhere from a few to a few hundred binaries disrupted by this disk phase, 
though our most detailed models suggest the number was $50$--$100$. 

First, we note that our fiducial model has a top-heavy initial mass function ($\alpha_m = 1.7$), motivated by observations of young stars in the central $0.5$ pc around Sgr A* \citep[e.g.,][]{Lu_2013}. The small dots show median estimates for $\alpha_m =1.5$ and $2.3$. At fixed NSC mass, smaller $\alpha_m$ values (flatter IMFs) yield fewer disruptions --- the average stellar mass is larger, so there are fewer stars in the NSC and therefore fewer binaries to disrupt.
Across this range of slopes, the number of disruptions varies by roughly a factor of four.

We show two other variations reflecting the uncertainty in cluster and disk properties. 
Making the cluster tangentially biased by setting $\beta = -0.3$ instead of $0$ reduces the number of disruptions by roughly a factor of three (triangles).
A disk less massive by a factor $\mu/\mu_{\rm fiducial}$ reduces the number of disruptions by a factor $(\mu/\mu_{\rm fiducial})^{1/2-\beta}$ --- see discussion around equation~(\ref{eq:prop_mu}) --- as shown by the estimates with $\mu=0.05$ instead of $0.1$ (xs).

Lastly, we show estimates accounting for two additional aspects of binary dynamics relevant before the arrival of the disk: 
\begin{enumerate}
    \item Our fiducial model assumed that \emph{all} binaries will be depleted over time, when in fact some binaries near the outskirts of the cluster are dynamically hard and should not evaporate. Therefore we make an estimate in which, prior to the disk phase, hard binaries are only depleted by reaching the end of their main sequence lifetimes (purple points). 
    This improved physics can cause a drastic increase in the number of disruptions from older populations (e.g., in the Chen et al.-based model).
    \item Our fiducial model neglected the idea that some fraction of binaries may \emph{become} hard through high-eccentricity oscillations and tidal friction. Such hardening may occur for $\sim 20\%$ of soft binaries at $a\sim r_h$ and a greater fraction at smaller $a$ \citep[see][]{Dodici_2026}. We make a final estimate (pink points) in which this shrinking is modelled stochastically (see Appendix~\ref{sec:appendix_stochastic}). This improved physics only leads to a small increase in the number of disruptions.
\end{enumerate}
Considering these aspects of pre-disk binary dynamics, we suggest that $10^2$ is an order-of-magnitude-accurate estimate of the number of disruptions from ADDD during the formation of the CWD.

Our estimates may be conservative in that 
we have only considered orbits with $a \leq r_h$. Outer-orbits just beyond the radius of influence will also be torqued by an accretion disk, increasing the number of disruptions by a factor of a few.
We do not include these binaries in our estimates, however, since the extent of the CWD-forming disk is unclear. 

In the following subsections, we use results from the fiducial model (i.e., the blue, circular points) to examine properties of S-stars and hypervelocity stars.

\subsection{S-star formation}

\begin{figure}
    \centering
    \includegraphics[width=\linewidth]{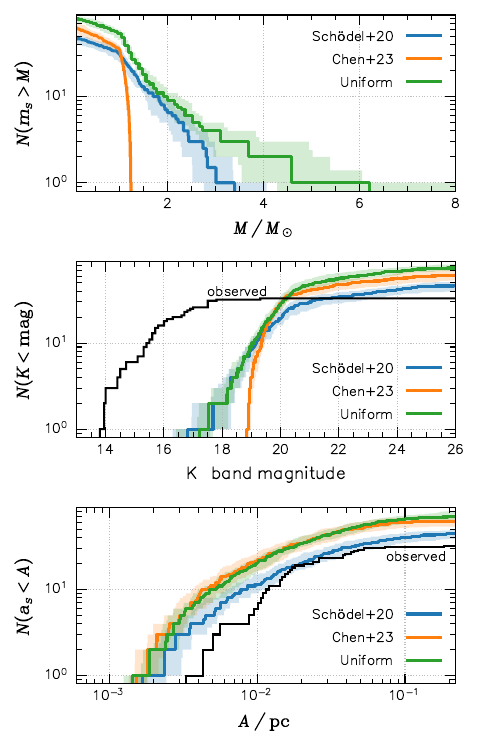}
    \caption{Properties of S-star analogues implanted by ADDD during the formation of the CWD, under three assumed NSC star-formation histories. We show the distributions of masses (top), K-band magnitudes (middle), and semimajor axes (bottom); the latter two panels also show distributions of detected S-stars \citep{Gillessen_2017}. Theoretical distributions are not corrected with selection functions or post-implantation dynamics. The ``turn-overs'' in the mass and magnitude distributions are set by the cutoff of the model IMF and, in reality, are likely not so notable.}
    \label{fig:s-stars}
\end{figure}

In Figure~\ref{fig:s-stars}, we show the masses ($m_s$), K band magnitudes, and semimajor axes ($a_s$), of the S-stars implanted by ADDD.

\subsubsection{Methods}
The semimajor axes are set to \citep[e.g.,][]{Sari_2010,Generozov_2020,Generozov_2025}
\begin{equation}
    a_s = 0.028 a_{\rm in} \left(\frac{m_b}{10\,M_\odot}\right)^{-2/3} \left(\frac{m_b}{m_b - m_s}\right).
\end{equation}
Eccentricities (not shown) are set such that an S-star retains the pericenter separation of its binary progenitor, which typically results in $e_s > 0.95$.

To estimate K band magnitudes, we use synthetic photometry for JWST NIRCam filter F210M from MIST v2.5 isochrones \citep{MIST1,MIST2,MIST3,MIST4} for Solar-metallicity stars rotating at $0.4\times$ breakup speed. We take the distance to the stars as 8.3 kpc \citep{GRAVITY_2021} and set the extinction to 2.42 mag \citep{Fritz_2011}.

\subsubsection{Results}

Most of the implanted stars are too dim for their orbits to have been characterized by VLT and Keck --- 
in each model, only a few to ten are as bright as S301, the dimmest S-star with a reported orbit \citep{S301}.
ELT/MICADO should improve the confusion limit by up to 7 mag relative to these 10-m-class telescopes \citep[e.g.,][]{vonFellenberg_2026}, such that we may determine properties of this low-mass tail of the S-stars. 

The mass function --- and thereby the magnitude function --- depends strongly on the pre-disk star formation history. The highest-mass stars implanted by ADDD come from the most-recent star formation epoch before the arrival of the disk. 
If this most-recent epoch was some time $t_{\rm delay}$ before the disk phase, then ADDD should not implant any stars with main sequence lifetimes shorter than $t_{\rm delay}$.
Explaining the brightest observed S-stars through ADDD would require a significant binary population formed in the NSC only a few Myr before the CWD-forming disk.
Lacking evidence for such a population, we presume that most S-stars implanted by ADDD during this disk phase have not yet been observed.

\subsubsection{Present-day orbits of low-mass S-stars}

Should these not-yet-observed S-stars share the orbital properties of the brighter, younger, observed sample?
In particular, should the eccentricity distribution be roughly thermal and the orbits isotropically oriented \citep[e.g.,][]{Gillessen_2017}? 

If the observed S-stars were also implanted by the Hills mechanism, within the last $5$ Myr (as suggested by their young age; \citealp{Habibi_2017}), then both their initial orbital properties and subsequent dynamical histories \citep[e.g.,][]{Bromley_2012,Lu_2021,Generozov_2025,Rom_2026} should be similar to the ADDD-implanted population proposed here. 
A natural conclusion would be that the both populations now have similar orbital properties.

That said, it is worth noting that ADDD implants S-stars early in the life of the gas disk, while the observed S-stars are typically thought to have been born in that disk and placed on short-period orbits at later times \citep[e.g.,][]{Generozov_2020,Verberne_2025}.
As a result, ADDD-implanted S-stars likely underwent many collisions with the gaseous disk ($10^4$--$10^6$, for a disk lifetime $\sim 10^6$ yr), while the younger S-stars did not. Collisions tend to diminish orbits' eccentricities and align them with the disk; in extreme cases, they may destroy a star altogether through mass loss or in-disk migration \citep[e.g.,][]{Artymowicz_1993,Macleod_2020,Fabj_2020,Generozov_2023}.
Efficient alignment with the disk could therefore impart the low-mass S-stars with some net sense of rotation, though resonant relaxation after the disk disappears may wipe away this signal. 

In Appendix~\ref{sec:appendix_stardisk}, we suggest that star-disk collisions may be influential in sculpting the orbital properties of ADDD-implanted S-stars, as well as S-stars implanted prior to the formation of the CWD. 
Considering the uncertainties inherent in such a discussion, though, we defer full analysis of this effect to future work.

\subsection{Hypervelocity star generation}

There is one HVS confidently associated with the Galactic Center and a Hills-mechanism origin: S5-HVS1 \citep{Koposov_2020}. This star was ejected $4.8$ Myr ago, consistent with ejection through ADDD during the formation of the CWD (for more, see Section~\ref{sec:S5-HVS1}).
Where might one find other HVSs ejected in the same burst? 

\subsubsection{Methods}

Using the package \texttt{speedystar} \citep{Evans_2022a}, we simulate $5$ Myr of ejected stars' flight under a static model of the Milky Way potential \citep{Bovy_2015}. Initial velocity vectors have nearly isotropic orientations (see below), set by the velocity vectors at disruption (i.e., at the first pericenter passage with $a(1-e) < r_t$) from our outer-orbit numerical integrations. Ejection speeds are \citep[e.g.,][]{Yu_2003,Rossi_2014}
\begin{equation}\label{eq:v_ej}
    v_{\rm ej} = \sqrt{\frac{2Gm_s}{a_{\rm in}}}\left(\frac{m_\bullet}{m_b}\right)^{1/6}.
\end{equation}

\begin{figure}
    \centering
    \includegraphics[width=\linewidth]{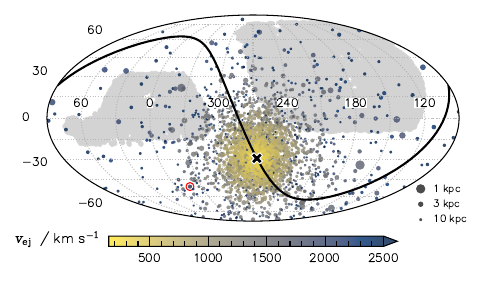} 
    \caption{On-sky positions, in ra and dec (degrees), of simulated HVSs ejected during the formation of the CWD (oversampled by a factor of 50 for visualization). Color and size indicate ejection speed and current heliocentric distance, respectively. S5-HVS1 is shown for reference, circled in red at $(344, -51)$. The black line and x show the Galactic midplane and Center, respectively. The grey region shows the DESI footprint.}
    \label{fig:on-sky}
\end{figure}

\subsubsection{Results}\label{sec:hvs_isotropy}

In Figure~\ref{fig:on-sky}, we show the present on-sky positions of these stars. Most HVSs are still near the Galactic midplane --- only $\approx 5\%$ have travelled far enough from Sgr A* to reach the DESI footprint, for example.

The velocity distribution of ejected stars may be estimated by combining equation~(\ref{eq:dist_result}), i.e., the distribution of disrupted-binary semimajor axes, with equation~(\ref{eq:v_ej}). This yields a comparable distribution of velocities to standard HVS literature \citep{Rossi_2014}, truncated at some minimum velocity (set by the widest disrupted binary).

\begin{figure}
    \centering
    \includegraphics[width=\linewidth]{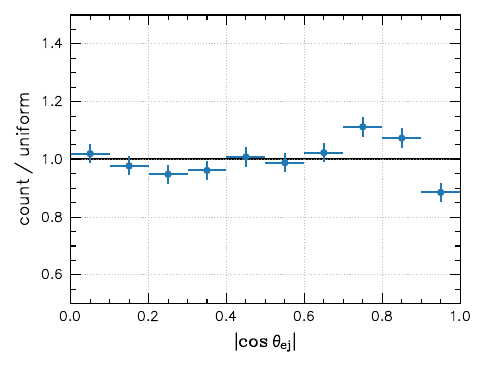}
    \caption{HVS ejections are only anisotropic at the $\sim 10\%$ level, which may be difficult to detect in observed distributions.  
    This figure shows the deviation from isotropy across a range of polar angles for velocity vectors at disruption among simulated outer orbits.
    Bin widths are shown by horizontal lines; vertical lines show Poisson errors.}
    \label{fig:isotropy}
\end{figure}

It is notable, and somewhat unintuitive, that ADDD yields roughly isotropic ejections.
In Figure~\ref{fig:isotropy}, we see that the distribution of polar angles among ejection velocity vectors only deviates from isotropy at the $\sim 10\%$ level; azimuthal angles are fully consistent with isotropy (not shown, but expected from the axisymmetry of the potential).

The features of the polar-angle distribution are set by the typical outer-orbital parameters upon disruption, which in turn are set by the evolution of orbits under a disk potential. The polar angle is determined by 
\begin{equation}
    |\cos \theta_{\rm ej}| = \sin i \cos \omega,
\end{equation}
where the orbital elements are those during the disrupting pericenter passage.
A typical disruption has $\ell_z \ll 1$ and starts with $e \ll 1$ --- requiring an initial $\sin i \sim 1$ --- and enters the loss cone while travelling along a contour in the librating island. 
Contours near the edge of the island (i.e, spanning the widest range of $\omega$) are most likely to reach the loss cone, as they come closest to $\ell = \ell_z$.

What is the relation between $i$ and $\omega$ for orbits on these contours?
When contours are nearly ``vertical'' (i.e., along constant $\omega$), orbits have $\sin i \lesssim 1$ and $\cos\omega \lesssim 1$; the product of these two values leads to the slight bump at $|\cos \theta_{\rm ej}| \sim 0.8$.
As these orbits reach small angular momenta, $\sin i$ values fall toward $0$; $\cos\omega$ values do not change drastically until the orbits almost reach their minimum $\ell$, at which point $\sin i$ is nearly minimized and $\omega$, suddenly, may span a wide range of values.

Orbits may enter the loss cone at any point on one of these contours, yielding a broad range of possible $\sin i \cos \omega$ --- and thereby a broad range of ejection angles. 
Orbits jumping into the loss cone rather than slowly evolving along contours (as noted in Section~\ref{sec:numerics}) further smooth this distribution.

This near-isotropy appears inconsistent with the findings of \citet{Penoyre_2025}, who argue that disruptions driven by long-lived axisymmetric potentials preferentially occur on near-polar orbits. 
We compare the two results in Appendix~\ref{sec:appendix_penoyre}. There, we conclude that the different distributions arise because their disruptions are dominated from binaries driven into the loss wedge, whereas ours are dominated by binaries already in it.

\section{Discussion}
\label{sec:discussion}

\subsection{Nuclear transients in other galaxies}\label{sec:transients}

So far we have focused on binaries undergoing ADDD in our own Galactic Center, as these yield byproducts that may be directly observed. ADDD should occur during an active phase of any galactic nucleus, though.
As a result, all AGN phases should increase the number of S-star analogues around a central MBH.

Some of these S-stars will have orbits bringing them close to the \emph{stellar} tidal disruption radius. 
Subsequent evolution --- two-body relaxation, disk-driven evolution, scalar resonant relaxation, etc.~(again, see \citealp{Lu_2021,Generozov_2025,Rom_2026}) --- may push pericenter separations below this radius, resulting in a TDE \citep{Hills_1975_tde,Rees_1988}. This pushing should be gradual, so the TDE may be partial at first, repeating on the orbital period of the S-star ($\sim 1$--$10$ yr). Repeating, partial TDEs have been observed \citep[e.g.,][]{Payne_2021} and may comprise a significant fraction of the TDE catalog \citep[e.g.,][]{Somalwar_2025,Yao_2026,Pan_2026}.

If delivery to the tidal disruption radius is common, these implanted S-stars may help explain the observed over-abundance of TDEs in post-starburst galaxies \citep[e.g.,][]{Arcavi_2014,French_2016,Hammerstein_2021}. In a starburst galaxy shut off by an AGN, ADDD likely produced an abundant population of S-stars that may subsequently be driven to tidal disruption.

In our sampled NSCs, the number of S-stars implanted by ADDD of binaries is smaller by at least a factor a few than the number of single stars driven directly to disruption by the disk.\footnote{While the loss cone for binary disruptions is somewhat larger, the small binary fraction among old clusters overcomes this increase.} 
At a glance, this imbalance suggests that the single stars are more important in setting the post-AGN TDE rate.   
However, the TDEs induced directly by the disk would occur while the disk is still growing \citepalias{Kaur_2025}, while S-stars evolving into TDEs would do so after some delay (see below), so these would have distinct observational epochs.
We further note that the single stars driven directly to disruption will rarely repeat on observable timescales (if they do repeat, most will have periods $\gg 100$ yr), whereas we expect S-stars driven to disruption after implantation to repeat on periods $\lesssim 10$ yr. 
The ratio of repeaters to non-repeaters among observed TDEs in post-starburst galaxies may help distinguish these two channels.

The delay between implantation of an S-star and its possible tidal disruption is uncertain. One might estimate this delay using the timescale for the angular momentum to be relaxed by of order itself via two-body interactions \citep[e.g.,][]{BT_2008}
\begin{align}
    t_{\rm 2BR} &\sim \frac{1}{\ln \Lambda} \left(\frac{m_\bullet}{m_\star}\right)^2 \frac{1-e_s}{N(< a_s)} P_s, 
\end{align}
where $P_s$ is the implanted star's period, $a_s$ is its semimajor axis, and $e_s$ is its eccentricity; $N(<a_s)$ is the number of stars internal to $a_s$.

This timescale is quite sensitive to $\gamma_c$, via $N(<a_s)$. For fiducial parameters ($\gamma_c = 5/4$), taking $a_s = 0.005$ pc and $e_s = 0.95$, we find $t_{\rm 2BR} \sim 10$ Gyr; however, changing $\gamma_c$ to $7/4$ --- which does not substantially change the outcome of ADDD --- shortens this to $t_{\rm 2BR} \sim 600$ Myr. 
The timescale is also sensitive to the presence of a compact-object cusp around the MBH, which could shorten $t_{\rm 2BR}$ both by increasing the number of bodies internal to $a_s$ and by increasing the average perturber mass $m_\star$.
The timescale becomes shorter, but remains uncertain, when one considers scalar resonant relaxation  \citep[see, e.g.,][]{Generozov_2025}. 

We defer full consideration of this channel of generating nuclear transients to future work (see also \citealp{Rom_2026}).

\subsection{Other processes generating HVSs}\label{sec:hvs_rates}

There are a host of previously known processes causing HVS ejections from the Galactic Center, both through Hills-mechanism disruptions (i.e., driving binaries into the loss cone) or through three-body interactions involving single stars. Here, we compare and contrast signatures of ADDD with several of these processes. We summarize this section in Table~\ref{tab:mechanisms}. 

Most work concerning HVSs presents an average rate of ejections. For ADDD, this rate is saddled with major uncertainties, as it requires knowledge not only of the number of disruptions per disk phase, but also of the recurrence cycle of massive disks. Standard estimates suggest that a typical MBH has had an accretion disk for $10^6$--$10^9$ yr of its life \citep[e.g.,][]{Yu_2002,Marconi_2004,Martini_2004} and that individual disk phases may persist for $10^5$--$10^7$ yr \citep[e.g.,][]{Schawinski_2015}, so a given nucleus may have experienced between $1$ and $10^4$ disk phases over a Hubble time. 
Even if we knew precisely the number of disruptions per disk phase, this range implies that the rate of ADDD, averaged over the age of a galaxy, cannot be presently constrained to uncertainties better than four orders of magnitude. 

Such uncertainties are typical of both theoretical and observational constraints on Hills-mechanism disruption rates. Observational bounds primarily rely on the detection of S5-HVS1 and a \emph{lack} of HVSs detected in Gaia DR3 and DESI data. Analyses generally exclude ejection rates $\lesssim 10^{-6}$ yr$^{-1}$ and $\gtrsim 10^{-3}$ yr$^{-1}$ (e.g., \citealp{Evans_2022a,Verberne_2025b}), though note that \citet{Verberne_2024} place a stronger constraint, excluding rates $\gtrsim 10^{-5}$ yr$^{-1}$. 
Most theoretical estimates --- including our estimate above, if we took $10^2$ disruptions per disk phase --- are consistent with $\lesssim 10^{-4}$ yr$^{-1}$; they are also often consistent with each other \citep[for a summary, see][]{Brown_2015}.

\begin{table}
    \centering
    \begin{tabular}{lccc}
        \toprule
        Process                      & Isotropy       & Burst      & Hills $v_{\rm ej}$ \\
        \midrule 
        Two-body relaxation        & \checkmark      &             & \checkmark  \\
        Aspherical Galactic Center   &      &              & \checkmark \\
        (I)MBH--MBH binary             &  $\sim$          & \checkmark  &   \\
        Eccentric-disk instabilities   &                 & \checkmark  & \checkmark  \\
        \hline
        \textbf{Accretion disk-driven} & $\sim$      & {\checkmark} & \checkmark  \\
        \bottomrule
    \end{tabular}
    \caption{
    ADDD is unique among known processes driving HVS ejections. It combines roughly isotropic ejection orientations (cf.~Figure~\ref{fig:isotropy}) with a bursty ejection rate and velocities set by the Hills mechanism. See text for references.}
    \label{tab:mechanisms}
\end{table}

The true power to distinguish ADDD may come from simultaneously considering the distributions of HVS ejection times, orientations, and velocities.
Here, we argue that ADDD is unique among HVS-inducing processes in that it drives Hills-mechanism ejections in temporal bursts --- of order the disk growth timescale \citepalias{Kaur_2025} --- with roughly isotropic orientations within each burst --- see Figure~\ref{fig:isotropy} and related discussion.

Several standard mechanisms provide a steady source of binary disruptions. These include the ``original'' method of loss-cone feeding: two-body relaxation, i.e., scattering interactions with background stars \citep[e.g.,][]{Lightman_1977,Yu_2003} or with massive perturbers like molecular clouds \citep[e.g.,][]{Perets_2007}. 
The non-spherical structure of the Galactic Centre --- on scales $\sim 1$--$10^{3}$ pc --- also falls into this category, driving binaries into the loss cone through a constant torque \citep{Magorrian_1999,Vasiliev_2013,Penoyre_2025}. 
The former yields isotropic ejections, while the latter preferentially ejects stars along the symmetry axis \citep[Section 6,][]{Penoyre_2025}.

Other mechanisms invoke transient phenomena, producing temporal bursts of HVS ejections. 
Another MBH or an intermediate-mass black hole (IMBH) merging with the MBH may cause HVS ejections; single stars in a galactic nucleus may be accelerated dramatically through interactions with such massive binaries 
(e.g., \citealp{Yu_2003,Gualandris_2005,Baumgardt_2006}).
If an eccentric disk of stars forms around the MBH, its self-gravity may drive binaries entrained within it onto loss cone orbits 
\citep{Madigan_2009,Haas_2016}. 
In these processes, ejections are primarily aligned with the plane of the (I)MBH orbit or disk, respectively. Ejections at late times during an IMBH inspiral may become isotropic \citep{Levin_2006}, but (i) anisotropy should still be present at the start of the burst and (ii) the distribution of ejection velocities from this process is distinctly biased toward slow $v_{\rm ej}$ relative to processes inducing Hills-mechanism ejections.

Therefore, to our knowledge, ADDD is unique among processes driving Hills-mechanism disruptions in that it features roughly isotropic ejections and bursty ejection rates. This argument is summarized in Table~\ref{tab:mechanisms}.

\subsection{The ejection of S5-HVS1}\label{sec:S5-HVS1}

S5-HVS1, ejected $\approx 5$ Myr ago at $\approx 1800$ km s$^{-1}$, is the fastest-known hypervelocity star and the one most-confidently associated with a Galactic Center origin \citep{Koposov_2020}. It was argued by \citet{Lu_2021} that --- assuming a constant ejection rate of $10^{-6}$ yr$^{-1}$ for stars between $2.5\,M_\odot$ and $4\,M_\odot$, and approximating that $\approx 20\%$ of ejected stars have velocities $> 10^3$ km s$^{-1}$ --- there should only be $\sim 1$ star of this mass ejected at such speed in the last $5$ Myr. It is therefore somewhat surprising that this star would happen to be detected by the S5 survey, which covered only $\sim 1\%$ of the sky. 
This line of reasoning encourages an explanation for S5-HVS1 involving an excess of disruptions $\sim 5$ Myr ago.

Our models naturally produce S5-HVS1 analogues. Three-of-four realizations of the NSC, using \citet{Schodel_2020}-based and uniform SFHs, yield at least one HVS matching the \citet{Lu_2021} criteria in both mass and speed. The typical realization had $\sim 10$ HVSs ejected at $> 1500$ km s$^{-1}$ --- a more conservative cut in speed, with no cut in mass.

Previous authors have suggested that S5-HVS1 may have come from a binary disruption driven by an eccentric-disk instability \citep[e.g.,][]{Generozov_2020b}. This explanation requires that the CWD formed as an eccentric disk and that the outer-orbital plane of the binary progenitor for S5-HVS1 coincidentally aligned with the disk plane. The disk then drove the binary onto a radial outer orbit, at which point it underwent a Hills-mechanism disruption.

We suggest that ADDD is a more natural explanation for S5-HVS1, primarily because it allows the middle-aged progenitor binary (the star itself appears to be $\approx 10^{7.7^{+0.25}_{-0.33}}$ yr old; \citealp{Koposov_2020}) to start in the larger phase-space area of orbits misaligned with the disk. It also does not require the somewhat fine-tuned requirement of an initially eccentric CWD. 
Of course, it is difficult to make a strong claim on the genesis of any particular HVS; further observations of HVSs ejected at similar times will help constrain the relative importance of these processes.

\subsection{Binaries that are not disrupted}

We have focused on ADDD, i.e., binaries whose outer orbits are torqued to the point of Hills-mechanism disruption by the potential of a massive accretion disk. While ADDD has clear observable implications, the majority of binaries in the cluster will \emph{not} become disrupted --- the loss wedge and librating island, together, cover a very small region of parameter space. 

Now, we briefly discuss how binaries that are not disrupted may be driven to reach very small \emph{inner}-orbit pericenter separations; stellar binaries may therefore shrink through tidal friction, while compact object binaries may merge through gravitational wave (GW) emission.

Binaries experience oscillations in $e_{\rm in}$ driven by the tidal potential of the MBH --- these are the classic von Zeipel-Lidov-Kozai (ZLK) cycles (\citealp{vonZiepel_1909,Lidov_1962,Kozai_1962}; see also \citealp{Naoz_2016,Tremaine_2023}).
The inner-orbit oscillations have a timescale
\begin{equation}
    t_{\rm ZLK} \sim  \frac{2}{3\pi} \left(\frac{a}{a_{\rm in}}\right)^{3/2}\left(\frac{m_b}{m_\bullet}\right)^{1/2}\left(1-e^2\right)^{3/2}P,
\end{equation}
with $P$ the outer-orbit period.
The outer-orbit precession from the cluster has a timescale
\begin{equation}\label{eq:t_sec}
    t_{\rm sec} \sim \left(\frac{a}{r_h}\right)^{\gamma_c-3}P,
\end{equation}
i.e., longer than the orbital period by the MBH mass over the enclosed cluster mass. 
The timescale for oscillation around the librating island is $\sim t_{\rm sec}\chi^{-1/2}$ (\citealp{Vasiliev_2013}; \citetalias{Kaur_2025}).

\begin{figure}
    \centering 
    \includegraphics[width=\linewidth]{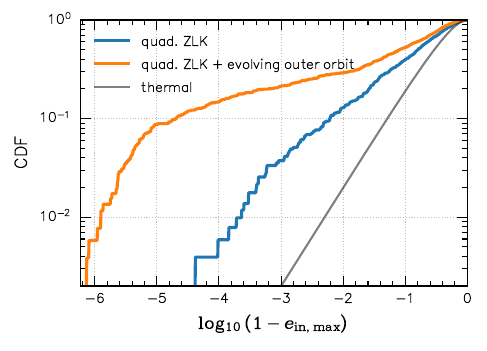}
    \caption{Secular chaos, driven by outer-orbit evolution, can cause binaries that do not become disrupted to reach very large inner-orbit eccentricities. Here, we compare the results of quadrupole-order ZLK cycles (blue) to binaries subject to the same equations of motion, but with their outer orbits torqued by a massive disk (orange). A thermal CDF is shown in grey, for reference.}
    \label{fig:inner_e}
\end{figure}
We illustrate the outcome of disk-induced secular chaos in Figure~\ref{fig:inner_e}. We show CDFs of one minus the maximum $e_{\rm in}$ for a set of binaries evolving under the quadrupole-order, ZLK equations of motion. In one case, the outer orbits are fixed; in another, outer orbits evolve under the fiducial disk and cluster potential. In the latter case, eccentricities reach values orders-of-magnitude closer to unity. 

When inner- and outer-orbit oscillation timescales are comparable, i.e., $t_{\rm sec}\chi^{-1/2} \sim t_{\rm ZLK}$, the binary is said to be in a regime of ``secular chaos'' \citep[e.g.,][]{Petrovich_2017,Bub2020}, where its inner-orbit variations are much less ordered than under ``standard'' ZLK cycles. Systems undergoing secular chaos are much more likely to reach some critically large $e_{\rm in}$. 

It has previously been suggested that high-$e_{\rm in}$ excursions by stellar binaries in galactic nuclei may lead to stellar mergers \citep[e.g.,][]{Prodan_2015,Stephan_2016,Stephan_2019}. Recent work, considering more-rigorous models of tidal friction, suggests that binaries should tend to shrink in $a_{\rm in}$ rather than merge through these excursions \citep{Dodici_2026,Huang_2026}.
Disk-induced secular chaos should increase the fraction of shrunken stellar binaries in a galactic nucleus. 

Compact object binaries driven to $1-e_{\rm in}\ll 1$ may begin to lose orbital binding energy via GW radiation \citep{Peters_1964}, leading to their inspiral and merger.
While works over the last decade have considered the hydrodynamical influence of massive accretion disks in driving GW mergers in galactic nuclei \citep[e.g.,][]{Bartos_2017, Stone_2017, Tagawa_2020, Ford_2022, McKernan_2025, Su_2025}, secular chaos induced by disks' potentials has been neglected.
This effect warrants further consideration.

\section{Conclusions}

We have examined the efficacy of massive accretion disks in driving Hills-mechanism disruptions of binary stars in galactic nuclei. We call this process accretion-disk-driven disruption (ADDD). 
ADDD occurs for binaries whose orbits around the MBH are (initially) nearly perpendicular to the disk plane; see Section~\ref{sec:orbits} for an overview of which orbits lead to disruptions.

We characterized the number of disruptions from an arbitrary disk in an arbitrary cluster (Section~\ref{sec:estimates}). Disruptions are more prevalent in radially biased clusters of stars formed a short time before a disk phase. Disks drive more disruptions when they are more massive and less centrally concentrated, but these factors are less influential than cluster properties. Figure~\ref{fig:num_vs_cluster} summarizes these results.

Through ADDD, the disk that formed the young stars in our Galactic Center may have produced observable S-stars and HVSs. If this disk was $\sim10 \times$ the mass of the observed CWD of young stars, we estimate it induced $\sim 10^2$ binary disruptions.
The S-stars and HVSs produced by ADDD would be \emph{older} than the young stellar population that comprises the CWD, so this mechanism struggles to explain the young, B-type stars observed in the S-star cluster.
The properties of these S-stars are shown in Figure~\ref{fig:s-stars}; the current on-sky distribution of these HVSs is shown in Figure~\ref{fig:on-sky}.

We note that ADDD may be a primary driver of HVS ejections in our Galactic Center (Section~\ref{sec:hvs_rates}) and that it may have been responsible for S5-HVS1 (Section~\ref{sec:S5-HVS1}). Its importance can be made clear with more detections of HVSs confidently associated with Galactic Center origin, through joint consideration of the distribution of ejection times, velocities, and orientations (see Table~\ref{tab:mechanisms}).

In other galaxies, ADDD may produce S-star analogues that are prone to subsequent tidal disruption, which may help explain the observed excess of TDEs in post-starburst galaxies. While we briefly discuss this process in Section~\ref{sec:transients}, we defer thorough consideration to future work.

We conclude by emphasizing that while the physics behind ADDD is robust, the predictions we can make regarding the number of disruptions and the properties of byproducts depend on highly uncertain characteristics of galactic nuclei. Primary sources of uncertainty are the velocity anisotropy, the star-formation history, and the initial mass function slope.
\vspace{0.3 cm}
 
\noindent
I first thank Cristobal Petrovich for asking me a question that broadly inspired this work. 
I am deeply grateful to Scott Tremaine and Yanqin Wu for their continually thoughtful advice and criticism. 
Many thanks are due to Fraser Evans, Zephyr Penoyre, Elena Maria Rossi, Mor Rozner, Biancamaria Sersante, Nicholas Stone, and Sebastiano von Fellenberg for helpful conversations.
I thank the Institute for Advanced Study for several months of hospitality, and I acknowledge support from
the Lachlan Gilchrist Fellowship and NSERC grants RGPIN-2020-03885 and RGPIN-2024-05533.

\bibliography{references}

\begin{appendix}

\section{Potentials}\label{sec:appendix_pots}

\subsection{Cluster}

Our spherical star cluster has density $\rho_c \propto r^{-\gamma_c}$, with a total mass $m_\bullet$ enclosed within $r_h$ --- we call this the radius of influence of the MBH (
though there are other definitions of this term).
For $\gamma_c \neq 2$, the cluster potential is
\begin{equation}
    \Phi_c = \frac{\Phi_{\bullet,h}}{2-\gamma_c}{\left(\frac{r}{r_h}\right)}^{2-\gamma_c},
\end{equation}
where $\Phi_{\bullet,h} \equiv -Gm_\bullet/r_h$ is the potential of the MBH at the radius of influence. 
The average of this potential over a Keplerian orbit is \citep[e.g., Section 4.4.1 of][]{Merritt_2013}
\begin{equation}
    \langle \Phi_c \rangle = \frac{\Phi_{\bullet,h}}{2-\gamma_c}{\left(\frac{a}{r_h}\right)}^{2-\gamma_c} f_c(e).
\end{equation}
The factor $f_{c}$ involves the hypergeometric function; we use the approximation (e.g., \citealp{Merritt_2013}; \citetalias{Kaur_2025})
\begin{align}\label{eq:fc_approx}
    f_{c}(e) &\approx 1 + \alpha e^2 + \mathcal{O}(e^4),
    \\ \label{eq:alpha}
    \text{with}\quad\quad \quad \alpha &\equiv \frac{2^{3-\gamma_c}}{\sqrt{\pi}}\frac{\Gamma(7/2 - \gamma_c)}{\Gamma(4-\gamma_c)} - 1,
\end{align}
where $\Gamma(x)$ is the gamma function.
For $\gamma_c \in [1,2]$, $\alpha$ declines monotonically from $0.5$ to $0$.

\subsection{Disk}
For the disk, we adopt the \citetalias{Kaur_2025} family of potential-density pairs. These pairs have two parameters: the radial power-law parameter $\gamma_d$ and the disk mass ratio $\mu \equiv m_d/m_\bullet$, with $m_d$ the disk mass within $r_h$. The potential is 
\begin{align}
    \Phi_d = 6\mu\Phi_{\bullet,h} \left(\frac{3-\gamma_d}{8-\tg}\right)
    {\left(\frac{r}{r_h}\right)}^{2-\gamma_d} T(\theta),
\end{align}
with $t_\gamma = (2-\gamma_d)(3-\gamma_d)$. The angular terms are
\begin{align} \nonumber 
    T(\theta) &= - \frac{(\tg - 2)(8-\tg)}{2\tg(6-\tg)} + |\cos\theta| + \frac{\tg-2}{2(6-\tg)}\cos^2\theta.
\end{align}
The disk density profile corresponding to this potential is presented in \citetalias{Kaur_2025}.

The average of this disk potential over a Keplerian orbit has the form 
\begin{equation}
   \langle \Phi_d \rangle = \mu \Phi_{\bullet,h} {\left(\frac{a}{r_h}\right)}^{2-\gamma_d}f_d(e,i,\omega).
\end{equation}
The factor $f_d$ is not analytical for general $\gamma_d$, but \citet{Kaur_2018} and \citetalias{Kaur_2025} derive the useful cases of $\gamma_d = 3/2$ and $5/2$. In this work, we numerically evaluate $f_d$ for a range of $\gamma_d$ in order to determine phase portraits (and in simulations, we use the non-averaged potential).  

One benefit of these potential-density pairs is that the ratio between scale height and radius, $h/r$, is constant. In standard AGN disk models, this ratio either is roughly constant or grows with $r$ in the outer regions \citep{Sirko_2003,Thompson_2005,Hopkins_2024}. Still, the \citetalias{Kaur_2025} models are a better approximation than standard finite-thickness disk potentials in galactic dynamics, in which $h$ itself is usually constant \citep[e.g.,][]{BT_2008}.
See Appendix A of \citetalias{Kaur_2025} for a full derivation of their disk model.

\subsection{Combined, orbit-averaged potential}

Although the orbit-averaged disk potential is generally not analytic,  
it is informative to write the combined, orbit-averaged potential of the cluster and disk (scaled by the MBH potential at $r_h$) in the form
\begin{equation}\label{eq:phi_p}
    \langle \phi \rangle \equiv \frac{\langle \Phi_c \rangle + \langle \Phi_d \rangle}{\Phi_{\bullet,h}} 
    = 
    f_c{\left(\frac{a}{r_h}\right)}^{2-\gamma_c} + \mu f_d {\left(\frac{a}{r_h}\right)}^{2-\gamma_d}.
\end{equation}
Plugging in equation~(\ref{eq:fc_approx}), dropping the resultant term that varies only with $a$, and scaling by $(a/r_h)^{\gamma_c-2}(2-\gamma_c)/\alpha$ yields a reduced form of the orbit-averaged potential \citepalias{Kaur_2025}:
\begin{equation}\label{eq:simple_phi}
    \langle {\varphi} \rangle = e^2 + \chi f_d(e,\omega,i;\gamma_d),
\end{equation}
where $\chi$, the strength of the disk potential relative to the cluster potential, is defined in equation~(\ref{eq:chi}). 
Note that when $\chi = 0$ --- i.e., the disk is not present --- the potential does not depend on $\omega$, so the angular momentum $\ell$ is conserved.

\subsection{Fits to separatrix action}

\begin{figure}
    \centering
    \includegraphics[width=\linewidth]{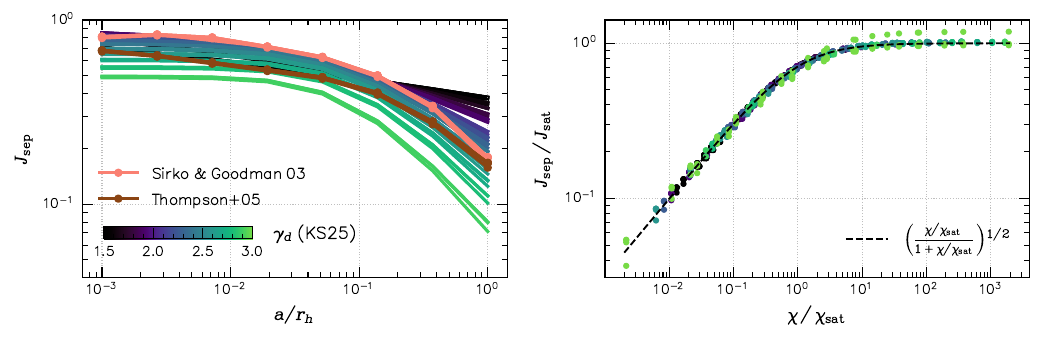}
    \caption{The action of the librating island separatrix is well-fit by equation~(\ref{eq:jhyp_fit}). In the left panel, we show numerically evaluted separatrix actions as a function of $a$ for \citetalias{Kaur_2025} disks with a range of physical properties. Color indicates $\gamma_d$ (dark purple is small, bright green is large). We also show the action for \citet{Sirko_2003} and \citet{Thompson_2005} disk models, which have similar shapes to \citetalias{Kaur_2025} disks with $2.5\lesssim \gamma_d <3$. In the right panel, we show the agreement between these results and the proposed fitting formula.}
    \label{fig:jsep_summary}
\end{figure}

In Table~\ref{tab:jsep_fits} and Figure~\ref{fig:jsep_summary}, we summarize our fits to the separatrix action formula given in equation~(\ref{eq:jhyp_fit}). For each $\gamma_d$ value tested, we numerically evaluate equation~(\ref{eq:simple_phi}) across $\ell \in [\ell_z,1]$ and $\omega \in [0,\pi]$ for a range of $a$ and $\ell_z \leq 0.01$, with $\gamma_c = 5/4$ or $7/4$ and $\mu = 0.01$ or $0.1$. We determine the separatrix by finding the minimum of $\langle {\varphi} \rangle$ with $\omega = 0$, which marks a hyperbolic fixed point, then integrating to find the area enclosed by this separatrix (i.e., $J_{\rm sep}$).

Given $\gamma_d$, each combination of $a$, $\gamma_c$, and $\mu$ yields a value of $\chi$. We conduct fits to equation~(\ref{eq:jhyp_fit}) using all combinations of these parameters for a given $\gamma_d$. Figure~\ref{fig:jsep_summary} suggests that these fits are generally precise. The worst fit comes from $\gamma_d = 2.9$ (bright green points in left panel), for which some values deviate significantly from the fit at $\chi \gg \chi_{\rm sat}$. 

\begin{table}[h]                          
      \centering                            
      \begin{tabular}{lcc}                  
          \hline
          $\gamma_d$ & $J_{\rm sat}$ & $\chi_{\rm sat}$ \\            
          \hline  
           1.5* & $1.095\pm 0.044$ & $1.71\pm 0.20$ \\          
           1.7 & $0.962\pm 0.002$ & $1.66\pm 0.01$ \\
           1.9 & $0.895\pm 0.010$ &  $2.07\pm 0.08$ \\          
           2.1 & $0.834\pm 0.009$ &  $2.40\pm 0.13$\\         
           2.3 & $0.769\pm 0.007$ &  $2.89\pm 0.14$ \\
           2.5 & $0.685\pm 0.005$ &  $3.58\pm 0.14$ \\         
           2.7 & $0.599\pm 0.009$ &  $4.72\pm 0.42$ \\
           2.9 & $0.501\pm 0.008$ &  $10.39\pm 0.94$ \\         
          \hline                            
      \end{tabular}                         
      \caption{Parameters for fits to equation~(\ref{eq:jhyp_fit}) for a range of disk concentration parameters $\gamma_d$. *The fit for $\gamma_d=1.5$ is poorly constrained because there are no data at $\chi \geq \chi_{\rm sat}$; this poor constraint leads to $J_{\rm sat} > 1$.}
      \label{tab:jsep_fits}                   
\end{table}

\subsection{Comparison to other disk models}

Finally, we compare properties of the librating island for the \citetalias{Kaur_2025}-family disks with standard literature models of AGN disks --- namely, \citet{Sirko_2003} and \citet{Thompson_2005}. 
We use \texttt{pagn} \citep{Gangardt_2024}, with all default parameters (but with $m_\bullet = 4\times10^{6}\,M_\odot$), to construct disk density models.

To approximate the potentials of these disks, we treat the density as a collection of massive rings, distributed uniformly in $\log_{10}(R)$ and in ${\rm arcsinh}(z)$ (the latter helps resolve the dense midplane of the disk). Mass is prescribed to these rings in a manner conserving the radial surface density profile, assuming the vertical density profile is Gaussian. At cylindrical position $(R,z)$, a ring of mass $M_{\rm ring}$, radius $R_{\rm ring}$, and height $z_{\rm ring}$ contributes a potential \citep[e.g.,][]{Bannikova_2011}
\begin{equation}
    \Phi_{\rm ring}(R,z) = 
    -\frac{GM_{\rm ring}}{\pi R_{\rm ring}} 
    {\left(\frac{R_{\rm ring}}{R}\right)}^{1/2} K\left(k^2\right)k, \quad \text{with} \quad 
    k^2 = \frac{4RR_{\rm ring}}{(R + R_{\rm ring})^2 + (z - z_{\rm ring})^2},
\end{equation}
where $K$ is the complete elliptic integral of the first kind. We create a lookup table of $\Phi_d/\mu$, as a function of position, for each model by summing over all rings, then use these lookup tables to create phase portraits for these disks.

The actions of the librating island separatrices are shown as a function of orbital semimajor axis in Figure~\ref{fig:jsep_summary}. Both the \citet{Sirko_2003} and \citet{Thompson_2005} models track the \citetalias{Kaur_2025} curves for $2.5\lesssim \gamma_d < 3$, the approximate slope of the density profiles of their outer regions.

\section{Approximate analytic solution of disrupted-binary distribution}
\label{sec:appendix_analytic}

We present an analytic approximation to the integral~(\ref{eq:integral_basic}).
Plugging in equations~(\ref{eq:DF}) and (\ref{eq:dF_da}), scaling by $N_c/N_h$ for a coeval cluster, then integrating over $\ell$ and $\ell_z$, the integral becomes
\begin{align}\label{eq:integ_1}
    \frac{1}{N_c}\left(\frac{dN_d}{da_{\rm in}}\right)_{m_b}
    &= \frac{2(1-\beta)(3-\gamma_{c})}{r_h(1-2\beta)} 
    J_{\rm sat}^{1-2\beta}
    {\left(\frac{df_{b,h}}{da_{\rm in}}\right)}
    {\left(\frac{2r_t}{r_h}\right)}^{1/2}
    \int_{r_m}^{r_h}da {\left(\frac{a}{r_h}\right)}^{3/2-\gamma_{c}+\lambda}  
    {\left(\frac{\chi/\chi_{\rm sat}}{1 + \chi/\chi_{\rm sat}}\right)}^{1/2-\beta},
\end{align}
where we have also used the fit to $J_{\rm sep}$ provided in equation~(\ref{eq:jhyp_fit}).
When $\gamma_d = \gamma_{c}$, the integrand simplifies to ${(a/r_h)}^{3/2-\gamma_{c}+\lambda}$, as $J_{\rm sep}$ is constant in $a$.
Otherwise, $\chi/\chi_{\rm sat} = {(a/r_{\rm sat})}^{\gamma_{c} - \gamma_d}$, where
$r_{\rm sat}$ is given by equation~(\ref{eq:rsat}).

The exact solution to equation~(\ref{eq:integ_1}) involves the hypergeometric function, but we can approximate it using a piecewise approximation to the term involving $\chi$ (see discussion around eq.~\ref{eq:jhyp_fit}).
This approximation overestimates the integrand by a factor $\sim 2$ at $a \approx r_{\rm sat}$, but is satisfactory elsewhere unless $|\gamma_d - \gamma_c| \ll 1$.

With this approximation, for $\gamma_{c} < \gamma_d$, 
\begin{align}
    \frac{1}{N_c}\left(\frac{dN_d}{da_{\rm in}}\right)_{m_b} &\approx 
    \frac{C}{r_h}
    {\left(\frac{df_{b,h}}{da_{\rm in}}\right)}
    {\left(\frac{r_t}{r_h}\right)}^{1/2}
    \left\{
    \int_{r_m}^{r_{\rm sat}}da {\left(\frac{a}{r_h}\right)}^{\Lambda}
    +
    \int_{r_{\rm sat}}^{r_h}da {\left(\frac{a}{r_h}\right)}^{\Lambda}
    {\left(\frac{a}{r_{\rm sat}}\right)}^{\Gamma}
    \right\}
\end{align}
with exponents $\Lambda \equiv 3/2 - \gamma_{c}+\lambda$ and $\Gamma \equiv (\gamma_{c} - \gamma_d)(1/2-\beta)$. The parameter $\Lambda$ encodes the radial profile of the binaries of interest, invoking both the cluster density profile and the binary fraction profile. The parameter $\Gamma$ encodes both the radial profile of the separatrix action and the anisotropy of the cluster. Typically, both of these are negative.

We have here defined the prefactor
\begin{equation}
    C(\gamma_{c}, \gamma_d, \beta) \equiv\frac{2\sqrt{2}(1-\beta)(3-\gamma_{c})}{(1-2\beta)} J_{\rm sat}^{1-2\beta}(\gamma_d).
\end{equation}
Then 
\begin{align} \label{eq:dN_dlna_case1}
     \frac{1}{N_c}\left(\frac{dN_d}{da_{\rm in}}\right)_{m_b} &\approx 
    C
    {\left(\frac{df_{b,h}}{d a_{\rm in}}\right)}
    {\left(\frac{r_t}{r_h}\right)}^{1/2}
    \left\{\frac{{(r_{\rm sat}/r_h)}^{\Lambda + 1} - {(r_m/r_h)}^{\Lambda + 1}}{\Lambda + 1}
    +
    \frac{1 - {(r_{\rm sat}/r_h)}^{\Lambda + \Gamma + 1}}{\Lambda + \Gamma + 1}
    {\left(\frac{r_h}{r_{\rm sat}}\right)}^{\Gamma}
    \right\}.
\end{align}
When $r_m > r_{\rm sat}$ (i.e., there are no binaries at $a$ where the separatrix is saturated), all instances of $r_{\rm sat}$ are replaced with $r_m$, except in the factor ${(r_h/r_{\rm sat})}^{\Gamma}$.

Similarly, when $\gamma_c > \gamma_d$, 
integration yields
\begin{align}\label{eq:dN_dlna_case2}
    \frac{1}{N_c}\left(\frac{dN_d}{da_{\rm in}}\right)_{m_b}
    &\approx 
    C
    {\left(\frac{df_{b,h}}{d a_{\rm in}}\right)}
    {\left(\frac{r_t}{r_h}\right)}^{1/2}
    \left\{\frac{1 - {(r_{\rm sat}/r_h)}^{\Lambda + 1}}{\Lambda + 1}
    +
    \frac{{(r_{\rm sat}/r_h)}^{\Lambda + \Gamma + 1} - {(r_m/r_h)}^{\Lambda + \Gamma + 1}}{\Lambda + \Gamma + 1}
    {\left(\frac{r_h}{r_{\rm sat}}\right)}^{\Gamma}
    \right\}.
\end{align}
When $r_{\rm sat} > r_h$ (i.e., there are no binaries at $a$ where the separatrix is saturated; this case is common), all instances of $r_{\rm sat}$ are replaced with $r_h$, except in the factor ${(r_h/r_{\rm sat})}^{\Gamma}$.

Equations~(\ref{eq:dN_dlna_case1}) and (\ref{eq:dN_dlna_case2}) are approximate analytic solutions to equation~(\ref{eq:integral_basic}). These expressions are generic, but somewhat opaque. We use the analytic calculation to help build understanding of the resultant distribution, but for most results we use Monte-Carlo solutions of the same integral.

\subsection{Agreement with Kaur \& Stone (2025)}

For $\gamma_d = 3/2$, equation~(\ref{eq:dN_dlna_case2}) is a natural extension of the \citetalias{Kaur_2025} result for the number of TDEs induced by an AGN disk potential. 
The orbits they consider are entirely in the $r_{\rm sat} > r_h$ regime, in which case equation~(\ref{eq:dN_dlna_case2}) simplifies to 
\begin{align}
    \frac{1}{N_c}\frac{dN_d}{d a_{\rm in}} 
    &\approx 
    C
    {\left(\frac{df_{b,h}}{da_{\rm in}}\right)}
    {\left(\frac{r_t}{r_h}\right)}^{1/2}
    {\left(\frac{r_h}{r_{\rm sat}}\right)}^{\Gamma}
    \left\{
    \frac{1 - {(r_m/r_h)}^{\Lambda + \Gamma + 1}}{\Lambda + \Gamma + 1}
    \right\}.
\end{align}
Since they consider single stars rather than binaries, $r_m \rightarrow 0$, and we integrate both sides over $a_{\rm in}$. Plugging in for $C$ and $r_{\rm sat}$ and setting $N_c = N_h$ yields
\begin{align}
    \frac{N_d}{N_h}
    &\approx 
    \frac{2\sqrt{2}(1-\beta)(3-\gamma_c)}{(1-2\beta)(\Lambda + \Gamma + 1)} J_{\rm sat}^{1-2\beta}
    {\left(\frac{r_t}{r_h}\right)}^{1/2}
    {\left[\frac{\mu}{\chi_{\rm sat}}\left(\frac{2-\gamma_c}{\alpha_\gamma}\right)\right]}^{1/2 - \beta}.
\end{align}
This result is equivalent to equation~(13) of \citetalias{Kaur_2025}. Their numerical prefactors are subsumed by our fitting parameters $J_{\rm sat}$ and $\chi_{\rm sat}$. 
From their Figure~5, their analytical estimates work best for $\gamma_c > \gamma_d$, which makes sense with the approximate form of $J_{\rm sep}$ they have used.  

\section{Stochastic model of binary shrinking}
\label{sec:appendix_stochastic}

\citet{Dodici_2026} find that a significant fraction of soft binaries orbiting MBHs should shrink to near-contact separations, through a combination of perturbed ZLK oscillations and dissipation of inner-orbital binding energy via dynamical stellar tides. When this occurs at $a \gtrsim 0.1$ pc around Sgr A*, for example, these shrunken binaries are dynamically hard, so they should be long-lived. 
We present a model of shrinking as a stochastic process, which recreates the fraction of binaries that shrink as a function of separation from Sgr A* \citep[Fig.~7]{Dodici_2026}.

The basis of this model is in the ZLK loss wedge discussed in that work. This wedge exists in a parameter space of values conserved under quadrupole-order ZLK oscillations. In the wedge, oscillations can bring the inner-orbit into the diffusive regime of dynamical tides \citep[e.g.,][]{Mardling_1995a,Ivanov_2004}, defined by a critical (inner-orbit) pericenter separation (eq.~1 of \citealp{Dodici_2026}). Binary--single interactions and vector resonant relaxation \citep[VRR;][]{Rauch_1996,Kocsis_2011,Kocsis_2015,Alexander_2017} cause systems to move around this parameter space. If a system is brought into the ZLK loss wedge --- and stays there for at least one full oscillation --- it will reach the diffusive-tide regime and shrink to near-contact separation.

A binary can only move around this parameter space for a limited time before it evaporates through interactions with background stars. We estimate that a binary's motion through the space is unbiased, so we approximate that there is a probability $p({\rm enter})=0.5$ that a binary enters the ZLK loss wedge before evaporation. If a binary was born in the ZLK loss wedge, we set $p({\rm enter}) = 1$.

Once in the wedge, it will not necessarily remain there for a full ZLK oscillation; it is more likely to do so if motion through the parameter space is slow relative to the oscillation period. If a binary enters the wedge, we say it will also reach the diffusive-tide regime and shrink with probability $p({\rm shrink}\,|\,{\rm enter}) = \left[1+10\left(t_{\rm ZLK}/t_{\rm evap}\right)\right]^{-1}$. The factor $10$ is the only fine-tuning of this model, set to match the empirical $p({\rm shrink}\,|\,{\rm enter})$ from the \citet{Dodici_2026} simulations. 

The probability of a soft binary shrinking is then $p({\rm enter})p({\rm shrink}\,|\,{\rm enter})$. We calculate this value for all sampled binaries and use it to weight a Boolean draw to determine whether or not a binary becomes hard through this process. 
If the draw says a binary should shrink, we say it does so after a time $t_{\rm evap} + t_{\rm ZLK}$ to account for the time it took to travel to the loss wedge. If the shrinking binary was born in the wedge, this time is instead set to $t_{\rm ZLK}$.

\section{Influence of star-disk collisions on S-star orbits}\label{sec:appendix_stardisk}

Comparing the surface density of a star to the surface density of the disk provides a simple estimate of the fractional change in the star's orbital properties per collision. 
If the disk density power law is maintained to very small $r$, for our fiducial parameters, an S-star with semimajor axis $a_s \sim 10^{-2}$ pc and $e_s = 0.98$ should encounter a disk surface density $\sim 10^6$ g cm$^{-2}$ at pericenter ($r \sim 10^{-4}$ pc).
The surface density of a Sun-like star is $M_{\odot} / \pi R_\odot^2 \sim 10^{11}$ g cm$^{-2}$. Comparing these values suggests significant orbital evolution --- or stellar mass loss --- may occur in $\sim 10^5$ orbits. 

S-stars have orbital periods of a few to $10^2$ years. If the disk lifetime is $\sim 10^6$ years, the gas disk may then have played a significant role in sculpting the orbital properties of ADDD-implanted S-stars.
Orbits that become aligned with the disk may still exhibit some level of alignment with each other, if they have only undergone a handful of vector resonant relaxation 
timescales.\footnote{Vector resonant relaxation reorients orbits through a torque that remains coherent for $t_{\rm coh} \sim 10^5$--$10^6$ yr in the S-cluster \citep[e.g.,][]{Kocsis_2011}. This torque acts identically on stars with identical orbits; therefore stars aligned with the gas disk should continue to have relatively similar orbit orientations through the first few $t_{\rm coh}$. After many $t_{\rm coh}$, small differences in initial conditions and subsequent response to the torques will accumulate, and orbit orientations will become isotropic. That said, particularly in the outer part of the S-cluster, there may not have been many $t_{\rm coh}$ since the formation of the CWD.}
Therefore it is possible that the low-mass S-stars 
will exhibit some net sense of rotation.
Eccentricities of these stars may be sub-thermal. 
In short, alignment with the gaseous disk would make it difficult to distinguish ADDD-implanted S-stars from low-mass stars born in that disk, without other probes of age.

Of course, star-disk interactions may not be as important as presented here. 
If our fiducial model overestimated the disk mass or the disk density power law parameter, the disk surface density (scaling like $\mu r^{1-\gamma_d}$) would also be overestimated. 
The same would be true if the power-law density profile is not maintained down to $r \lesssim 10^{-4}$ pc \citep[see, e.g., Figure 2 of][]{Sirko_2003}. 
Furthermore, as eccentricity decays, growth of the pericenter separation would lead to collisions with a less-dense part of the disk --- in this way, orbit evolution through star-disk collisions is somewhat self-limiting.
Each of these considerations makes star-disk collisions less consequential.

We defer rigorous modelling of the present-day properties of low-mass S-stars, considering interactions with the CWD-forming gas disk, to future work. Such modelling is particularly timely with the promise of ELT/MICADO observations of these stars in the next decade.

\section{Comparing near-isotropy of HVS ejections to recent work}\label{sec:appendix_penoyre}

Recently, \citet{Penoyre_2025} found that the axisymmetric (non-spherical) potential of the Galactic Center on wider scales drives disruptions that occur preferentially with $|\cos i|\sim 0$, i.e., while the orbit normal remains significantly misaligned from the symmetry axis.
Our results show a distribution of orbits at disruption that is much closer to isotropic than what they report (Section~\ref{sec:hvs_isotropy}). In fact, the distribution of $|\cos i|$ underlying our results has a slight preference for values $> 0.5$, and a \emph{deficit} at $|\cos i|\sim 0$.

This apparent discrepancy has a relatively straightforward explanation. 
We are interested in the probability density function of $\cos i$ for orbits at the moment of disruption.
This function --- for fixed $\gamma_z \equiv \ell_{z}/\ell_{\rm crit}$ --- is given in equation~(51) of \citet{Penoyre_2025}:\footnote{This equation assumes the probability density function $p(\beta) \propto \beta^{-3/2}$, with the diving factor $\beta = \left(\ell_{\rm crit}/\ell\right)^2$ at disruption. In our simulations, this distribution is somewhat steeper, i.e., more systems are disrupted at smaller $\beta$, because our orbits enter the loss cone on regular trajectories rather than chaotic ones. We neglect this difference for the sake of simplicity, but note that the argument in this appendix holds for steeper distributions of $\beta$.} 
\begin{equation}
    p(\cos i; \, \gamma_z) = \frac{|\gamma_z|}{1 - |\gamma_z|} \frac{1}{\cos^2 i},\quad \quad |\gamma_z| < |\cos i| < 1.
\end{equation}
This distribution is always peaked at small $\cos i$, so \emph{at a given} $\gamma_z$, most disruptions occur with the smallest possible $\cos i$.
The \emph{overall} distribution of $\cos i$ is found by marginalizing over $\gamma_z$ for the binaries that become disrupted.

In \citet{Penoyre_2025}, the binary disruption rate is set by the refilling rate of the loss wedge. They find that this refilling is fastest at very small $\gamma_z$, such that the distribution of $\gamma_z$ at disruption is peaked toward $0$. 
On the other hand, the distribution of $\gamma_z$ among our disruptions reflects the initial distribution of $\ell_z$ at small values. It does \emph{not} depend on the rate of refilling of the loss wedge. While it still declines with $\gamma_z$, the distribution of $\gamma_z$ is closer to flat.

\begin{figure}
    \centering
    \includegraphics[width=0.5\linewidth]{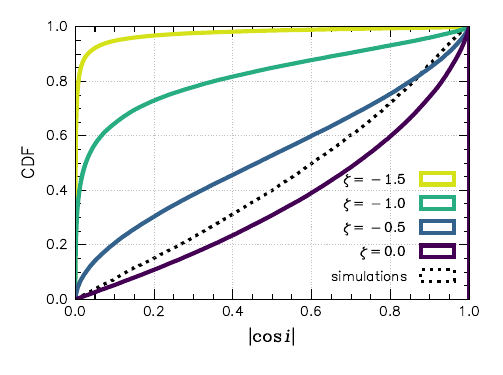}
    \caption{Cumulative distributions of $|\cos i|$ at disruption for a mock population of binaries under a range of possible distributions of $\gamma_z \equiv \ell_z / \ell_{\rm crit}$ (colored lines) and for the binaries disrupted in our simulations (black, dotted line).}
    \label{fig:cosi_dists}
\end{figure}

Say that $p(\gamma_z) \propto \gamma_z^\zeta$. Marginalizing over this distribution to find the overall distribution $p(\cos i)$ does not yield an analytic result for generic $\zeta$, so we instead create mock disrupted populations by sampling $10^5$ values from $p(\gamma_z)$, then sampling from $p(\cos i;\,\gamma_z)$ for each. 
Figure~\ref{fig:cosi_dists} shows the resultant $|\cos i|$ distributions for different values of $\zeta$. 

Distributions of $\gamma_z$ that are peaked toward $0$ (i.e., $\zeta < 0$) lead to distributions of $|\cos i|$ with a similar peak. Such peaks are not present in our simulations, nor in the mock population with $\zeta = 0$.

We conclude that the different distributions of $|\cos i|$ between our work and \citet{Penoyre_2025} primarily reflect different distributions of $\gamma_z$ among our disrupted binaries. In turn, these distributions reflect the different processes underlying the two studies. In their work, a static loss wedge must be refilled. In ours, a transient disk arises and presents a brand new loss wedge; disruptions come from orbits that were already filling that wedge, so no refilling is required.

\end{appendix}

\end{document}